\documentclass[a4paper, 11pt]{article}
\pdfoutput=1

\usepackage{jheppub}
\usepackage[utf8]{inputenc}
\usepackage[T1]{fontenc}
\usepackage{afterpage}
\usepackage{multirow}
\usepackage{rotating}

\newcommand{\Dmq}{\Delta m^2}
\newcommand{\dCP}{\delta_\text{CP}}
\newcommand{\Nuc}[2]{{\ensuremath{\mbox{}^{#1}}\text{#2}}}
\newcommand{\eVq}{\ensuremath{\text{eV}^2}}

\renewcommand{\Im}{\mathop{\mathrm{Im}}}
\newcommand{\diag}{\mathop{\mathrm{diag}}}
\newcommand{\SumNu}{{\textstyle\sum} m_\nu}
\newcommand{\NEW}{$\Rightarrow$}

\newenvironment{pagefigure}{\begin{figure}[!p]}{\afterpage\clearpage\end{figure}}

\makeatletter
\DeclareRobustCommand\recite[1]{\begingroup\@fileswfalse\cite{#1}\endgroup}
\makeatother

\graphicspath{{Figures/}}

\allowdisplaybreaks

\title{NuFit-6.0: Updated global analysis of three-flavor neutrino
  oscillations}

\author[a,b]{Ivan Esteban,}
\affiliation[a]{Department of Physics, University of the Basque
  Country UPV/EHU, PO Box 644, 48080 Bilbao, Spain}
\affiliation[b]{EHU Quantum Center, University of the Basque Country
  UPV/EHU, PO Box 644, 48080 Bilbao, Spain}
\emailAdd{ivan.esteban@ehu.eus}

\author[c,d,e]{M.~C.~Gonzalez-Garcia,}
\affiliation[c]{Departament de F\'{\i}sica Qu\`antica i
  Astrof\'{\i}sica and Institut de Ciencies del Cosmos, Universitat de
  Barcelona, Diagonal 647, E-08028 Barcelona, Spain}
\affiliation[d]{Instituci\'o Catalana de Recerca i Estudis
  Avan\c{c}ats (ICREA), Pg.\ Lluis Companys 23, 08010 Barcelona,
  Spain.}
\affiliation[e]{C.N.~Yang Institute for Theoretical Physics, State
  University of New York at Stony Brook, Stony Brook, NY 11794-3840,
  USA}
\emailAdd{concha.gonzalez-garcia@stonybrook.edu}

\author[f]{Michele Maltoni,}
\affiliation[f]{Instituto de F\'{\i}sica Te\'orica UAM/CSIC, Calle de
  Nicol\'as Cabrera 13--15, Universidad Aut\'onoma de Madrid,
  Cantoblanco, E-28049 Madrid, Spain}
\emailAdd{michele.maltoni@csic.es}

\author[g]{Ivan Martinez-Soler,}
\affiliation[g]{Institute for Particle Physics Phenomenology, Durham
  University, South Road, DH1 3LE, Durham, UK}
\emailAdd{ivan.j.martinez-soler@durham.ac.uk}

\author[c]{Jo\~ao Paulo Pinheiro,}
\emailAdd{joaopaulo.pinheiro@fqa.ub.edu}

\author[h]{Thomas Schwetz}
\affiliation[h]{Institut f\"ur Astroteilchenphysik, Karlsruher
  Institut für Technologie (KIT), 76021 Karlsruhe, Germany}
\emailAdd{schwetz@kit.edu}

\abstract{We present an updated global analysis of neutrino
  oscillation data as of September 2024.  The parameters
  $\theta_{12}$, $\theta_{13}$, $\Dmq_{21}$, and $|\Dmq_{3\ell}|$
  ($\ell = 1,2$) are well-determined with relative precision at
  $3\sigma$ of about 13\%, 8\%, 15\%, and 6\%, respectively.  The
  third mixing angle $\theta_{23}$ still suffers from the octant
  ambiguity, with no clear indication of whether it is larger or
  smaller than $45^\circ$.  The determination of the leptonic CP phase
  $\dCP$ depends on the neutrino mass ordering: for normal ordering
  the global fit is consistent with CP conservation within $1\sigma$,
  whereas for inverted ordering CP-violating values of $\dCP$ around
  $270^\circ$ are favored against CP conservation at more than
  $3.6\sigma$.  While the present data has in principle
  $2.5$--$3\sigma$ sensitivity to the neutrino mass ordering, there
  are different tendencies in the global data that reduce the
  discrimination power: T2K and NOvA appearance data individually
  favor normal ordering, but they are more consistent with each other
  for inverted ordering.  Conversely, the joint determination of
  $|\Dmq_{3\ell}|$ from global disappearance data prefers normal
  ordering.  Altogether, the global fit including long-baseline,
  reactor and IceCube atmospheric data results into an almost equally
  good fit for both orderings.  Only when the $\chi^2$ table for
  atmospheric neutrino data from Super-Kamiokande is added to our
  $\chi^2$, the global fit prefers normal ordering with $\Delta\chi^2
  = 6.1$.  We provide also updated ranges and correlations for the
  effective parameters sensitive to the absolute neutrino mass from
  $\beta$-decay, neutrinoless double-beta decay, and cosmology.}

\preprint{IFT-UAM/CSIC-24-140, YITP-SB-2024-24, IPPP/24/64}


\begin{document}

\maketitle

\section{Introduction}

The global analysis of neutrino oscillation data provides us with the
unique comprehensive description of the non-trivial flavor structure
of leptons.  Although in the last years some efforts are being put
forward by the experimental collaborations for combined analysis of
their results~\cite{T2K:2024wfn, T2K:nu24, NOvA:nu24}, the main task
still falls on the work of phenomenological
groups~\cite{Esteban:2020cvm, deSalas:2020pgw, Capozzi:2018ubv}.  Over
the last decade these global analyses have provided consistent results
with very good quantitative agreement on the dominant effects.  In a
nutshell, it is found that mass-driven oscillations between three
neutrino states of different mass and three different mixing angles
account for the vast majority of the results.  And, when redundant,
the results from different experiments on the dominant effects in this
picture are statistically compatible.

In addition, three-neutrino oscillation effects which are subdominant
or quantitatively small in present experiments include the mass
ordering (MO) of the three states, the possible maximality of one of
the mixing angles ($\theta_{23}$ in the standard parametrization), and
the amount of leptonic CP violation.  They remain open questions in
the present experimental landscape and constitute the main goal of the
upcoming generation of experiments~\cite{JUNO:2024jaw, DUNE:2020jqi,
  Hyper-Kamiokande:2018ofw}.  The phenomenological analysis finds
minor differences in these subdominant effects, and their statistical
significance has been varying with time as the running experiments
accumulate more statistics and update their analyses.  This, in fact,
is one of the main motivations for the redundancy of the global
phenomenological analysis, providing unbiased tests for the
consistency over time of the emerging picture.

In this effort, this work contains the latest analysis within the
NuFIT program, NuFIT~6.0, which incorporates a number of changes and
updates since our last published analysis in
2020~\cite{Esteban:2020cvm}.  In the solar neutrino sector, the new
generation of Standard Solar Models~\cite{B23Fluxes} are now used for
the predictions, and the full day-night spectrum from the phase-IV of
Super-Kamiokande~\cite{Super-Kamiokande:2023jbt} together with the
final spectra from Borexino phases-II~\cite{Borexino:2017rsf} and
III~\cite{BOREXINO:2022abl} are included.  For the long baseline (LBL)
reactor data, we have updated the reactor antineutrino fluxes used for
the predictions to the latest Daya-Bay
measurements~\cite{DayaBay:2021dqj}, and results from
SNO+~\cite{SNO:2024wzq, SNO+:nu24, SNO+poster:nu24} are included in
the global analysis for the first time.  From reactors at medium
baseline (MBL), we include the most up-to-date Daya-Bay spectral
data~\cite{DayaBay:2022orm}.  Updates in the LBL accelerator analysis
include incorporating the latest samples and simulation updates of
T2K~\cite{T2K:nu24}, and the doubled statistics of NOvA neutrino
samples~\cite{NOvA:nu24}.  A new independent analysis of atmospheric
neutrinos from 3 years of IceCube/DeepCore data~\cite{IceCube:2019dyb,
  IceCube:2019dqi} has been incorporated.  Finally, the effect of the
updated $\chi^2$ maps provided by the collaborations for the analysis
of the latest atmospheric neutrino samples in
Super-Kamiokande~\cite{Super-Kamiokande:2023ahc, SKatm:data2024} and
IceCube/DeepCore~\cite{IceCube:2024xjj, IC:data2024} is accounted for.

The outline of the paper is as follows.  In Sec.~\ref{sec:global24} we
present the results of our global analysis, providing best-fit values
and 1-dimensional and 2-dimensional confidence regions for the 6
oscillation parameters, and we discuss the global determination of
leptonic CP violation.  In Sec.~\ref{sec:atm} we discuss in some
detail the various tendencies in the global data, focusing on
experiments sensitive to the large mass splitting $\Dmq_{3\ell}$
($\ell = 1,2$).  In Sec.~\ref{sec:lbl} we discuss the compatibility
among T2K and NOvA appearance data, whereas in Sec.~\ref{sec:reacatm}
we consider the global combination of disappearance data (which
includes accelerator, reactor and atmospheric neutrinos), presenting a
detailed study of the compatibility among different combinations of
datasets.  In Sec.~\ref{sec:MO} we discuss in detail the sensitivity
to the neutrino mass ordering in terms of a proper hypothesis test.
Section~\ref{sec:solar} focuses on the ``solar sector'' governed by
$\Dmq_{21}$, especially in light of the latest developments in solar
models.  In Sec.~\ref{sec:absmass} we provide the relevant
correlations between absolute neutrino mass observables.  We summarize
our results in Section~\ref{sec:conclu}, and provide a list of all the
used data in Appendix~\ref{sec:appendix-data24}.  In
Appendix~\ref{sec:app_IC} we describe our analysis of IceCube data,
and in Appendix~\ref{sec:app-T} we provide more details on the mass
ordering test.

\section{Global analysis}
\label{sec:global24}

We start by presenting the results of the NuFIT~6.0 global fit.
Parametrization conventions and technical details on our global
analysis can be found in Ref.~\cite{Esteban:2018azc}.  In particular,
we use the standard parametrization of the $ 3\times 3$ unitary
leptonic mixing matrix~\cite{Maki:1962mu, Kobayashi:1973fv}
\begin{equation}
  \label{eq:U3m}
  U =
  \begin{pmatrix}
    1 & 0 & 0 \\
    0 & c_{23}  & {s_{23}} \\
    0 & -s_{23} & {c_{23}}
  \end{pmatrix}
  \cdot
  \begin{pmatrix}
    c_{13} & 0 & s_{13} e^{-i\dCP} \\
    0 & 1 & 0 \\
    -s_{13} e^{i\dCP} & 0 & c_{13}
  \end{pmatrix}
  \cdot
  \begin{pmatrix}
    c_{12} & s_{12} & 0 \\
    -s_{12} & c_{12} & 0 \\
    0 & 0 & 1
  \end{pmatrix}
   \cdot \mathcal{P} \,,
\end{equation}
where $c_{ij} \equiv \cos\theta_{ij}$ and $s_{ij} \equiv
\sin\theta_{ij}$, with angles $\theta_{ij} \in [0, \pi/2]$ and phase
$\dCP \in [0, 2\pi]$, such that $\dCP \neq 0,\pi$ implies CP violation
in neutrino oscillations in vacuum~\cite{Cabibbo:1977nk,
  Bilenky:1980cx, Barger:1980jm}.  Here $\mathcal{P} = I$ for Dirac
neutrinos and $\mathcal{P} = \diag(e^{i\alpha_1}, e^{i\alpha_2},1)$
for Majorana neutrinos, a matrix that plays no role in neutrino
oscillations~\cite{Bilenky:1980cx, Langacker:1986jv}.  In this
convention, there are two non-equivalent orderings for the neutrino
masses, namely normal ordering (NO) with $m_1 < m_2 < m_3$, and
inverted ordering (IO) with $m_3 < m_1 < m_2$.  In particular, in what
follows we use the definition
\begin{equation}
  \Dmq_{3\ell}
  \quad \text{with}\quad
  \begin{cases}
    \ell = 1 & \text{for $\Dmq_{3\ell} > 0$: normal ordering (NO),} \\
    \ell = 2 & \text{for $\Dmq_{3\ell} < 0$: inverted ordering (IO).}
  \end{cases}
\end{equation}

The analysis includes all data available up to September 2024 which,
for convenience, we list in Appendix~\ref{sec:appendix-data24} with
the corresponding references.  As is customary in the NuFIT analysis
since v4.0~\cite{Esteban:2018azc}, we show two versions of the
analysis.  These versions differ in the inclusion of atmospheric
neutrino results, for which there is not enough information for us to
make an independent analysis comparable to that performed by the
collaborations.  In NuFIT~6.0, this is the case for the atmospheric
neutrino data from Super-Kamiokande phases 1-5 (SK-atm) and from the
latest 9.3-year result from IceCube/DeepCore (IC24).  For those, we
use their tabulated $\chi^2$ maps provided in
Refs.~\cite{SKatm:data2024} and~\cite{IC:data2024}, respectively,
which we can combine with our global analysis for the solar, reactor
and LBL experiments.  We note that for IceCube/DeepCore we have
performed an independent analysis of their previous 3-year atmospheric
neutrino data sample~\cite{IceCube:2019dyb, IceCube:2019dqi}, which we
do include in the version of the analysis without tabulated $\chi^2$
maps.  In what follows, we refer as <<IC19 w/o SK-atm>> to our independent
analysis variant without tabulated $\chi^2$ maps, and as <<IC24 with
SK-atm>> to the comprehensive analysis variant that includes the
tabulated SK-atm and IC24 $\chi^2$ maps instead of our 3-year IceCube/DeepCore
analysis.

\begin{pagefigure}\centering
 \includegraphics[width=0.86\textwidth]{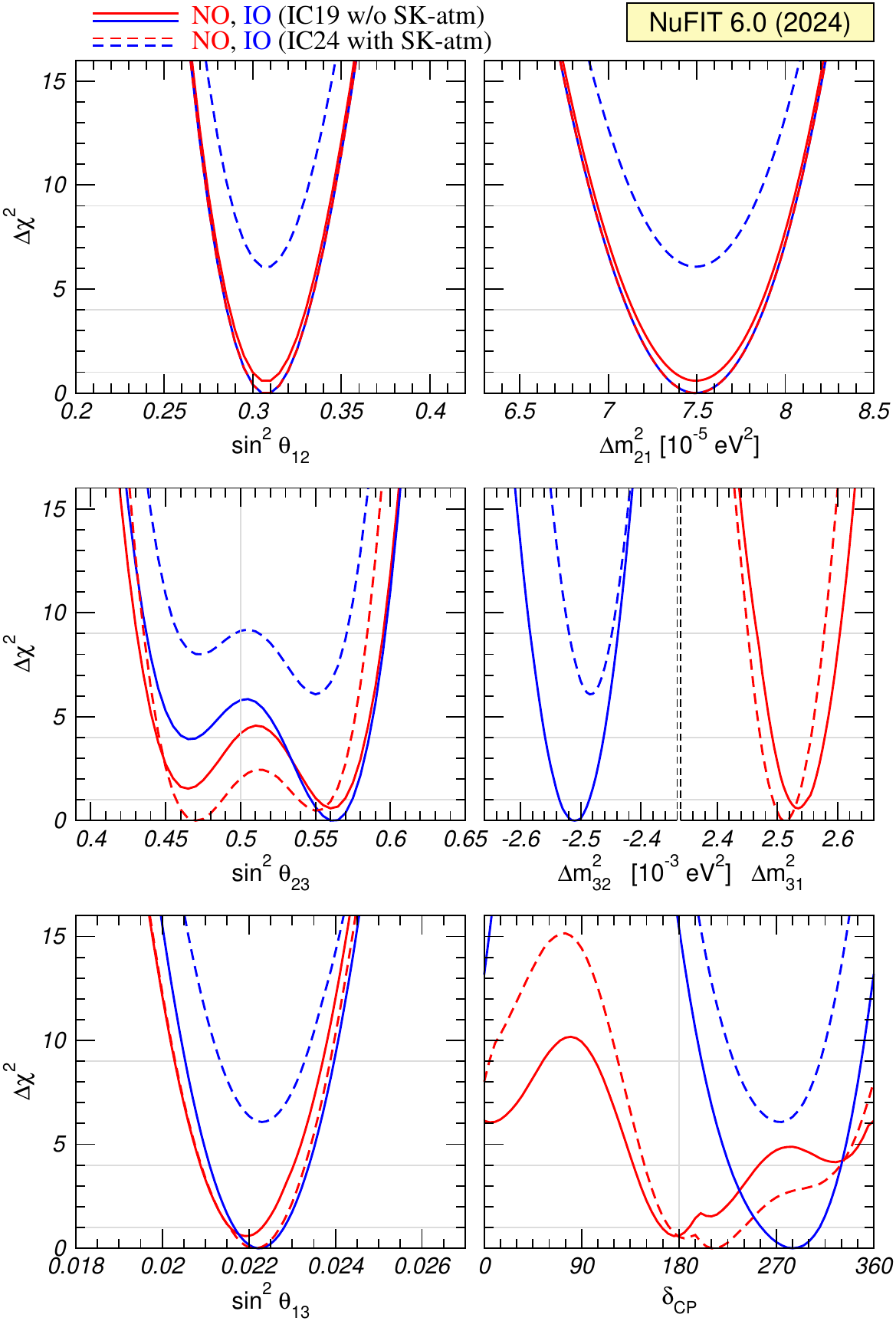}
  \caption{Global $3\nu$ oscillation analysis.  We show $\Delta\chi^2$
    profiles minimized with respect to all undisplayed parameters.
    The red (blue) curves correspond to Normal (Inverted) Ordering.
    Solid and dashed curves correspond to the two variants of the
    analysis as described in the labels (notice that in the upper two
    panels the red-dashed and blue-solid lines almost completely
    overlap). }
  \label{fig:chisq-glob24}
\end{pagefigure}

\begin{pagefigure}\centering
 \includegraphics[width=0.81\textwidth]{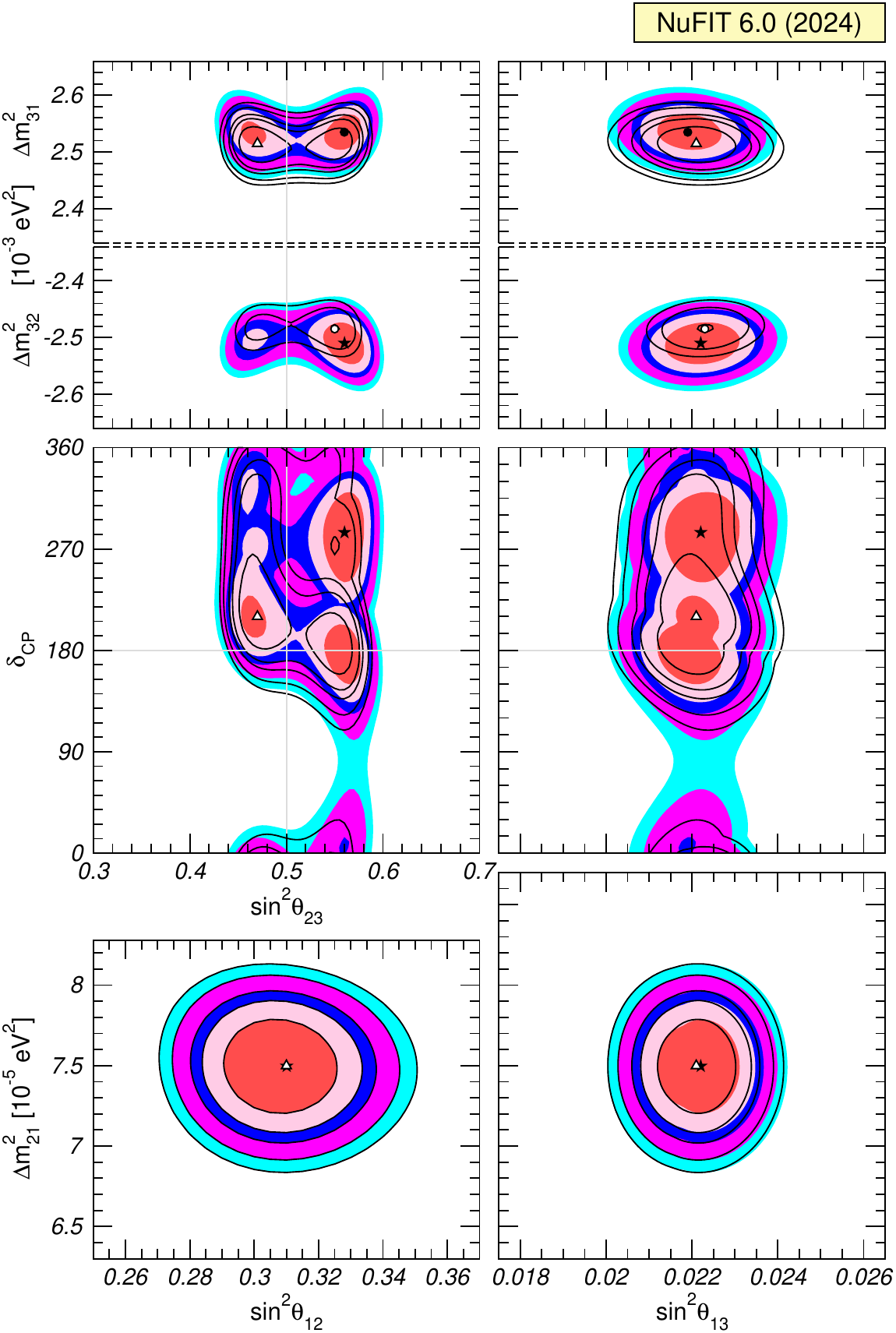}
  \caption{Global $3\nu$ oscillation analysis.  Each panel shows the
    two-dimensional projection of the allowed six-dimensional region
    after minimization with respect to the undisplayed parameters.
    The regions in the four lower panels are obtained from
    $\Delta\chi^2$ minimized with respect to the mass ordering.  The
    different contours correspond to $1\sigma$, 90\%, $2\sigma$, 99\%,
    $3\sigma$ CL (2 dof).  Colored regions (black contours) correspond
    to the variant with IC19 and without SK-atm (with IC24 and with
    SK-atm).}
  \label{fig:region-glob24}
\end{pagefigure}

A selection of the results of our global fit are displayed in
Fig.~\ref{fig:chisq-glob24} (one-dimensional $\Delta\chi^2$ curves)
and Fig.~\ref{fig:region-glob24} (two-dimensional projections of
confidence regions).  In Table~\ref{tab:bfranges24} we give the
best-fit values as well as $1\sigma$ and $3\sigma$ confidence
intervals for the oscillation parameters in both mass orderings,
relative to the local best-fit points in each ordering.  Additional
figures and tables corresponding to this global analysis can be found
in the NuFIT webpage~\cite{nufit}.

\begin{table}\centering
  \begin{footnotesize}
    \begin{tabular}{c|l|cc|cc}
      \hline\hline
      \multirow{11}{*}{\begin{sideways}\hspace*{-7em}IC19 without SK atmospheric data\end{sideways}} &
      & \multicolumn{2}{c|}{Normal Ordering ($\Delta\chi^2 = 0.6$)}
      & \multicolumn{2}{c}{Inverted Ordering (best fit)}
      \\
      \cline{3-6}
      && bfp $\pm 1\sigma$ & $3\sigma$ range
      & bfp $\pm 1\sigma$ & $3\sigma$ range
      \\
      \cline{2-6}
      \rule{0pt}{4mm}\ignorespaces
      & $\sin^2\theta_{12}$
      & $0.307_{-0.011}^{+0.012}$ & $0.275 \to 0.345$
      & $0.308_{-0.011}^{+0.012}$ & $0.275 \to 0.345$
      \\[1mm]
      & $\theta_{12}/^\circ$
      & $33.68_{-0.70}^{+0.73}$ & $31.63 \to 35.95$
      & $33.68_{-0.70}^{+0.73}$ & $31.63 \to 35.95$
      \\[3mm]
      & $\sin^2\theta_{23}$
      & $0.561_{-0.015}^{+0.012}$ & $0.430 \to 0.596$
      & $0.562_{-0.015}^{+0.012}$ & $0.437 \to 0.597$
      \\[1mm]
      & $\theta_{23}/^\circ$
      & $48.5_{-0.9}^{+0.7}$ & $41.0 \to 50.5$
      & $48.6_{-0.9}^{+0.7}$ & $41.4 \to 50.6$
      \\[3mm]
      & $\sin^2\theta_{13}$
      & $0.02195_{-0.00058}^{+0.00054}$ & $0.02023 \to 0.02376$
      & $0.02224_{-0.00057}^{+0.00056}$ & $0.02053 \to 0.02397$
      \\[1mm]
      & $\theta_{13}/^\circ$
      & $8.52_{-0.11}^{+0.11}$ & $8.18 \to 8.87$
      & $8.58_{-0.11}^{+0.11}$ & $8.24 \to 8.91$
      \\[3mm]
      & $\dCP/^\circ$
      & $177_{-20}^{+19}$ & $96 \to 422$
      & $285_{-28}^{+25}$ & $201 \to 348$
      \\[3mm]
      & $\dfrac{\Dmq_{21}}{10^{-5}~\eVq}$
      & $7.49_{-0.19}^{+0.19}$ & $6.92 \to 8.05$
      & $7.49_{-0.19}^{+0.19}$ & $6.92 \to 8.05$
      \\[3mm]
      & $\dfrac{\Dmq_{3\ell}}{10^{-3}~\eVq}$
      & $+2.534_{-0.023}^{+0.025}$ & $+2.463 \to +2.606$
      & $-2.510_{-0.025}^{+0.024}$ & $-2.584 \to -2.438$
      \\[2mm]
      \hline\hline
      \multirow{11}*{\begin{sideways}\hspace*{-7em}IC24 with SK atmospheric data\end{sideways}} &
      & \multicolumn{2}{c|}{Normal Ordering (best fit)}
      & \multicolumn{2}{c}{Inverted Ordering ($\Delta\chi^2 = 6.1$)}
      \\
      \cline{3-6}
      && bfp $\pm 1\sigma$ & $3\sigma$ range
      & bfp $\pm 1\sigma$ & $3\sigma$ range
      \\
      \cline{2-6}
      \rule{0pt}{4mm}\ignorespaces
      & $\sin^2\theta_{12}$
      & $0.308_{-0.011}^{+0.012}$ & $0.275 \to 0.345$
      & $0.308_{-0.011}^{+0.012}$ & $0.275 \to 0.345$
      \\[1mm]
      & $\theta_{12}/^\circ$
      & $33.68_{-0.70}^{+0.73}$ & $31.63 \to 35.95$
      & $33.68_{-0.70}^{+0.73}$ & $31.63 \to 35.95$
      \\[3mm]
      & $\sin^2\theta_{23}$
      & $0.470_{-0.013}^{+0.017}$ & $0.435 \to 0.585$
      & $0.550_{-0.015}^{+0.012}$ & $0.440 \to 0.584$
      \\[1mm]
      & $\theta_{23}/^\circ$
      & $43.3_{-0.8}^{+1.0}$ & $41.3 \to 49.9$
      & $47.9_{-0.9}^{+0.7}$ & $41.5 \to 49.8$
      \\[3mm]
      & $\sin^2\theta_{13}$
      & $0.02215_{-0.00058}^{+0.00056}$ & $0.02030 \to 0.02388$
      & $0.02231_{-0.00056}^{+0.00056}$ & $0.02060 \to 0.02409$
      \\[1mm]
      & $\theta_{13}/^\circ$
      & $8.56_{-0.11}^{+0.11}$ & $8.19 \to 8.89$
      & $8.59_{-0.11}^{+0.11}$ & $8.25 \to 8.93$
      \\[3mm]
      & $\dCP/^\circ$
      & $212_{-41}^{+26}$ & $124 \to 364$
      & $274_{-25}^{+22}$ & $201 \to 335$
      \\[3mm]
      & $\dfrac{\Dmq_{21}}{10^{-5}~\eVq}$
      & $7.49_{-0.19}^{+0.19}$ & $6.92 \to 8.05$
      & $7.49_{-0.19}^{+0.19}$ & $6.92 \to 8.05$
      \\[3mm]
      & $\dfrac{\Dmq_{3\ell}}{10^{-3}~\eVq}$
      & $+2.513_{-0.019}^{+0.021}$ & $+2.451 \to +2.578$
      & $-2.484_{-0.020}^{+0.020}$ & $-2.547 \to -2.421$
      \\[2mm]
      \hline\hline
    \end{tabular}
  \end{footnotesize}
  \caption{Three-flavor oscillation parameters from our fit to global
    data for the two variants of the analysis described in the text.
    The numbers in the 1st (2nd) column are obtained assuming NO (IO),
    \textit{i.e.}, relative to the respective local minimum.  Note
    that $\Dmq_{3\ell} \equiv \Dmq_{31} > 0$ for NO and $\Dmq_{3\ell}
    \equiv \Dmq_{32} < 0$ for IO.}
  \label{tab:bfranges24}
\end{table}

With these results, we obtain the following $3\sigma$ relative
precision of each parameter $x$, defined as $2(x^\text{up} -
x^\text{low}) / (x^\text{up} + x^\text{low})$, where $x^\text{up}$
($x^\text{low}$) is the upper (lower) bound on $x$ at the $3\sigma$
level:
\begin{equation}
  \label{eq:precision24}
  \begin{aligned}
    \theta_{12} &: 13\% \,,
    &\quad
    \theta_{13} &: \left\{
    \begin{array}{lr}
      \text{NO} & 8.1\% ~[8.2\%] \,, \\
      \text{IO} & 7.8\% ~[7.9\%] \,,
    \end{array}\right.
    &\quad
    \theta_{23} &: \left\{
    \begin{array}{lr}
      \text{NO} & 21\% ~[19\%] \,, \\
      \text{IO} & 20\% ~[18\%] \,,
    \end{array}\right.
    \\
    \Dmq_{21} &: 15\% \,,
    &\quad
    |\Dmq_{3\ell}| &: \left\{
    \begin{array}{lr}
      \text{NO} & 5.6\% ~[5.1\%] \,, \\
      \text{IO} & 5.8\% ~[5.1\%] \,,
    \end{array}\right.
    &\quad
    \dCP &: \left\{
    \begin{array}{lr}
      \text{NO} & 100\% ~[98\%] \,, \\
      \text{IO} & 54\% ~[55\%] \,,
    \end{array}\right.
  \end{aligned}
\end{equation}
where the numbers between brackets show the impact of including IC24
and SK-atm.  We note that given the non-Gaussianity of
$\Delta\chi^2(\dCP)$, the above estimated precision for $\dCP$ can
only be taken as indicative, in particular for NO.

Altogether, we derive the following $3\sigma$ ranges on the magnitude
of the elements of the leptonic mixing matrix (see
Ref.~\cite{GonzalezGarcia:2003qf} for details on how we derive the
ranges and their correlations):
\begin{equation}
  \label{eq:umatrix24}
  \begin{aligned}
    |U|_{3\sigma}^\text{IC19 w/o SK-atm} &=
    \begin{pmatrix}
      0.801 \to 0.842 &\qquad
      0.519 \to 0.580 &\qquad
      0.142 \to 0.155
      \\
      0.248 \to 0.505 &\qquad
      0.473 \to 0.682 &\qquad
      0.649 \to 0.764
      \\
      0.270 \to 0.521 &\qquad
      0.483 \to 0.690 &\qquad
      0.628 \to 0.746
    \end{pmatrix}
    \\[1mm]
    |U|_{3\sigma}^\text{IC24 with SK-atm} &=
    \begin{pmatrix}
      0.801 \to 0.842 &\qquad
      0.519 \to 0.580 &\qquad
      0.142 \to 0.155
      \\
      0.252 \to 0.501 &\qquad
      0.496 \to 0.680 &\qquad
      0.652 \to 0.756
      \\
      0.276 \to 0.518 &\qquad
      0.485 \to 0.673 &\qquad
      0.637 \to 0.743
    \end{pmatrix}
  \end{aligned}
\end{equation}

\begin{figure}\centering
  \includegraphics[width=0.9\textwidth]{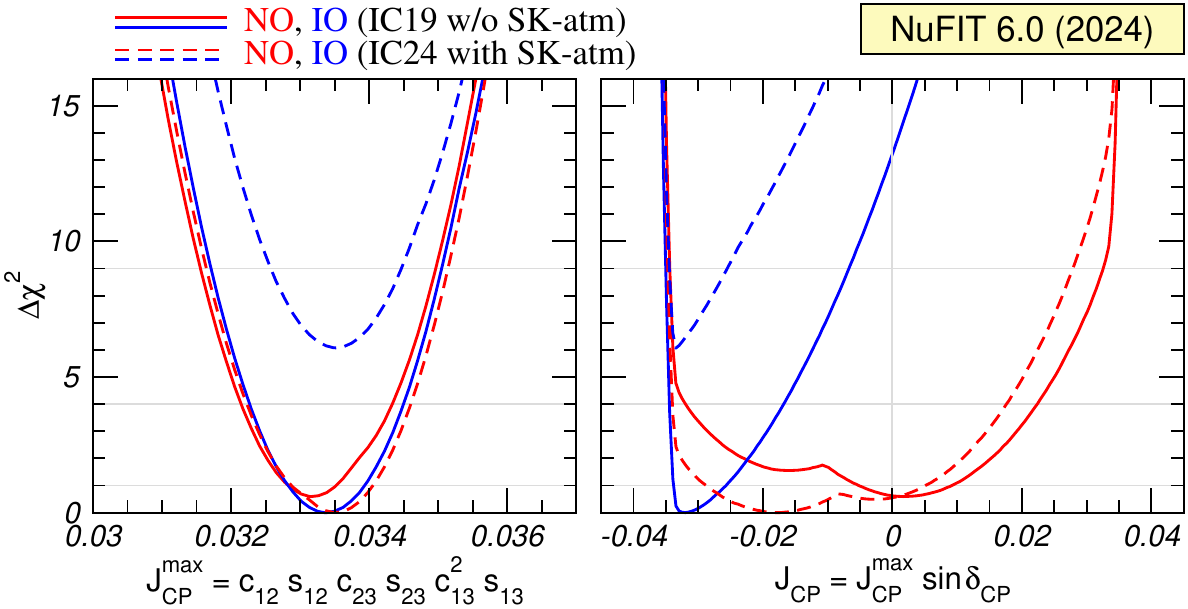}
  \caption{Dependence of the global $\Delta\chi^2$ function on the
    Jarlskog invariant.  The red (blue) curves are for NO (IO).  Solid
    (dashed) curves are for the <<IC19 w/o SK-atm>> (<<IC24 with
    SK-atm>>) $\Delta\chi^2$.}
  \label{fig:chisq-viola}
\end{figure}

We quantify the presence of leptonic CP violation in neutrino
propagation in vacuum in a convention-independent form in terms of the
leptonic Jarslkog invariant~\cite{Krastev:1988yu}:
\begin{equation}
  \label{eq:jcplep}
  \begin{split}
    J_\text{CP}
    &\equiv \Im\big[ U_{\alpha i} U_{\alpha j}^* U_{\beta i}^* U_{\beta j} \big]
    \\
    &\equiv J_\text{CP}^\text{max} \sin\dCP =
    \cos\theta_{12} \sin\theta_{12}
    \cos\theta_{23} \sin\theta_{23} \cos^2\theta_{13} \sin\theta_{13}
    \sin\dCP \,.
  \end{split}
\end{equation}
Its present determination is shown in Fig.~\ref{fig:chisq-viola}, from
which we read its maximum value
\begin{equation}
  \label{eq:jmax}
  J_\text{CP}^\text{max} = 0.0333 \pm 0.0007 \, (\pm 0.0017) \,.
\end{equation}
at $1\sigma$ ($3\sigma$) for both orderings.  $J_\text{CP}$ is totally
analogous to the invariant introduced in Ref.~\cite{Jarlskog:1985ht}
for the description of CP-violating effects in the quark sector,
presently determined to be $J_\text{CP}^\text{quarks} =
(3.12^{+0.13}_{-0.12}) \times 10^{-5}$~\cite{PDG}.

Figure~\ref{fig:chisq-viola} also shows that in NO the best-fit value
$J_\text{CP}^\text{best} = 0.0017\,(-0.018)$ (where the value in
parenthesis corresponds to the analysis with IC24 and SK-atm) is only
favored over CP conservation $J_\text{CP} = 0$ with
$\Delta\chi^2=0.02\,(0.55)$.  In contrast, in IO CP conservation is
disfavored with respect to $J_\text{CP}^\text{best} = -0.032$ with
$\Delta\chi^2=13\,(16)$, which corresponds to $3.6\sigma$ ($4\sigma$)
when evaluated for 1~dof.

\section{Status of neutrino mass ordering, leptonic CP violation, and \texorpdfstring{$\boldsymbol{\theta_{23}}$}{theta23}}
\label{sec:atm}

\subsection{Updates from T2K and NOvA}
\label{sec:lbl}

\begin{figure}\centering
  \includegraphics[width=0.85\textwidth]{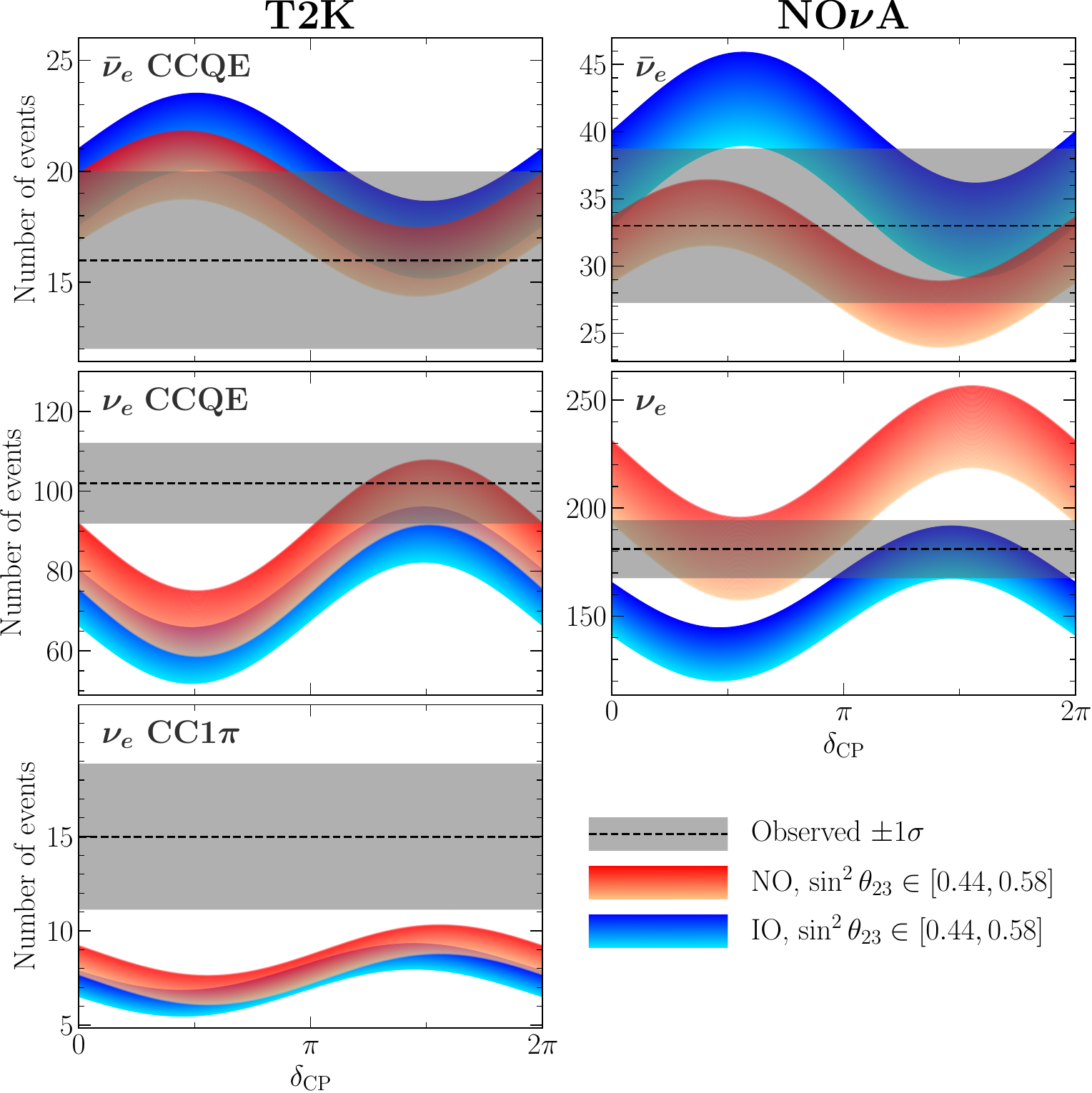}
  \caption{Predicted number of events as a function of $\dCP$ for the
    T2K (left) and NOvA (right) appearance data sets.
    $\sin^2\theta_{23}$ varies between 0.44 and 0.58, where the
    lower-light (upper-dark) bound of the colored bands corresponds to
    0.44 (0.58).  Red (blue) bands correspond to NO (IO).  For the
    other oscillation parameters we have adopted $\sin^2\theta_{13} =
    0.0222$, $|\Dmq_{3\ell}| = 2.5\times 10^{-3}~\eVq$,
    $\sin^2\theta_{12} = 0.32$, $\Dmq_{21} = 7.5\times 10^{-5}~\eVq$.
    The horizontal dashed lines show the observed number of events,
    with the $\pm 1\sigma$ statistical error indicated by the gray
    shaded band.}
  \label{fig:nevts24}
\end{figure}

We start by discussing the implications of the latest data from the
T2K and NOvA long-baseline accelerator experiments, presented at the
Neutrino 2024 conference.  Although our analysis includes the full
spectral information provided by the collaborations, qualitative
understanding of the results can be obtained from the study of the
total number of events in the different appearance samples.  To this
end, we show in Fig.~\ref{fig:nevts24} the predicted number of events
for these samples as a function of $\dCP$, for varying values of
$\sin^2\theta_{23}$, as well as the mass ordering, compared to the
observations.

The predictions in Fig.~\ref{fig:nevts24} are calculated using our
simulations of the experiments, that include numerically-computed
oscillation probabilities.  However, the general behaviour of the
curves is well-described by the approximate expressions derived in
Refs.~\cite{Elevant:2015ska, Esteban:2018azc}.  These expand the
relevant oscillation probabilities in the small parameters
$\sin\theta_{13}$, $\Dmq_{21}L/E_\nu$, and $A \equiv | 2E_\nu V /
\Dmq_{3\ell}|$ (where $L$ is the baseline, $E_\nu$ the neutrino energy
and $V$ the effective matter potential~\cite{Wolfenstein:1977ue}),
resulting in the following expressions for the expected number of
events:
\begin{align}
  N_{\nu_e}
  &\approx \mathcal{N}_\nu
  \left[ 2 s_{23}^2(1+2oA) - C' \sin\dCP(1+oA) \right] \,,
  \label{eq:Nnu}
  \\
  N_{\bar\nu_e} &
  \approx \mathcal{N}_{\bar\nu}
  \left[ 2 s_{23}^2(1-2oA) + C' \sin\dCP(1-oA) \right] \,,
  \label{eq:Nan}
\end{align}
where $o \equiv \text{sgn}(\Dmq_{3\ell})$.  For T2K the mean neutrino
energy gives $A \approx 0.05$, whereas for NOvA we find that the
approximation works best with the \emph{empirical} value of
$A=0.1$.  Furthermore, taking all of the well-determined parameters
$\theta_{13}$, $\theta_{12}$, $\Dmq_{21}$, $|\Dmq_{3\ell}|$ at their
global best-fit points, we obtain numerically $C' \approx 0.28$ with
negligible dependence on $\theta_{23}$.  The normalization constants
$\mathcal{N}_{\nu,\bar\nu}$ calculated from our re-analysis of T2K and
NOvA are given for the various appearance samples in
Table~\ref{tab:app24}, along with the corresponding observed event
numbers and expected backgrounds.  We can obtain insight on the results
of the fit by considering the ratio $r \equiv
(N_\text{obs}-N_\text{bck})/\mathcal{N}$.

\begin{table}\centering
  \catcode`?=\active\def?{\hphantom{0}}
  \catcode`!=\active\def!{\hphantom{.}}
  \begin{tabular}{c|cccccc}
    \hline\hline
    & \multicolumn{3}{c}{T2K ($\nu$)} & T2K ($\bar\nu$) & NOvA ($\nu$) & NOvA ($\bar\nu$) \\
    & CCQE & CC1$\pi$ & Sum & \\
    \hline
    $\mathcal{N}$              & ?54!? & ?5!? & ?59!? & 13 & 104!? & 23 \\
    $N_\text{obs}$              & 102!? & 15!? & 117!? & 16 & 181!? & 33 \\
    $N_\text{obs}-N_\text{bck}$  & ?81.6 & 12.5 & ?94.2 & 10 & 117.3 & 19 \\
    $r=\frac{N_\text{obs}-N_\text{bck}}{\mathcal{N}}$ & 1.5 (1.6) & 2.5 (2.4) & 1.59 (1.65) & 0.77 (0.61)& 1.13 (1.14) & 0.83 (0.83) \\
    \hline\hline
  \end{tabular}
  \caption{Normalization coefficients $\mathcal{N}_\nu$ and
    $\mathcal{N}_{\bar\nu}$ in eqs.~\eqref{eq:Nnu} and~\eqref{eq:Nan}
    for approximations used to qualitatively describe the appearance
    event samples for T2K and NOvA.  Numbers in parentheses are the
    corresponding values for the data set used for NuFIT~5.0.}
    \label{tab:app24}
\end{table}

From the numbers in the table we observe the following:
\begin{itemize}
\item T2K data has $r > 1$ for neutrinos and $r < 1$ for
  antineutrinos, so the square-bracket in Eq.~\eqref{eq:Nnu} has to be
  enhanced and the one in Eq.~\eqref{eq:Nan} suppressed.  This can be
  achieved with $\dCP \simeq 3\pi/2$, with a better fit in NO.  This
  preference has been present since the first T2K results on $\nu_e$
  appearance.  As can be seen in the last line of
  Table~\ref{tab:app24}, although the preference is somewhat weaker in
  NuFIT~6.0 than it was in NuFIT~5.0, it is still significant; and the
  strongest observed effect remains in the lower-statistics CC$1\pi$
  sample.

\item NOvA antineutrino results have not changed since NuFIT~5.0.
  They have $r<1$ and can be accommodated with either NO and $\dCP
  \simeq \pi/2$, or IO and $\dCP \simeq 3\pi/2$, with a slightly
  better fit in NO.

\item NOvA neutrino results have now 60\% more statistics than in
  NuFIT~5.0, and they still result in $r\sim 1$.  This can also be
  accommodated with either NO and $\dCP \simeq \pi/2$, or IO and $\dCP
  \simeq 3\pi/2$, totally compatible with the NOvA antineutrino
  results.  Altogether, the results from NOvA only show a very mild
  preference for NO.

\item The favored values of $\dCP$ in NO by T2K and NOvA do not agree.
  Consequently, the combination of both experiments is better
  described in IO with $\dCP \simeq 3\pi/2$.  This was already the
  case in NuFIT~5.0, and the tendency has strengthened with the
  updated results.
\end{itemize}

\begin{figure}\centering
  \includegraphics[width=0.7\textwidth]{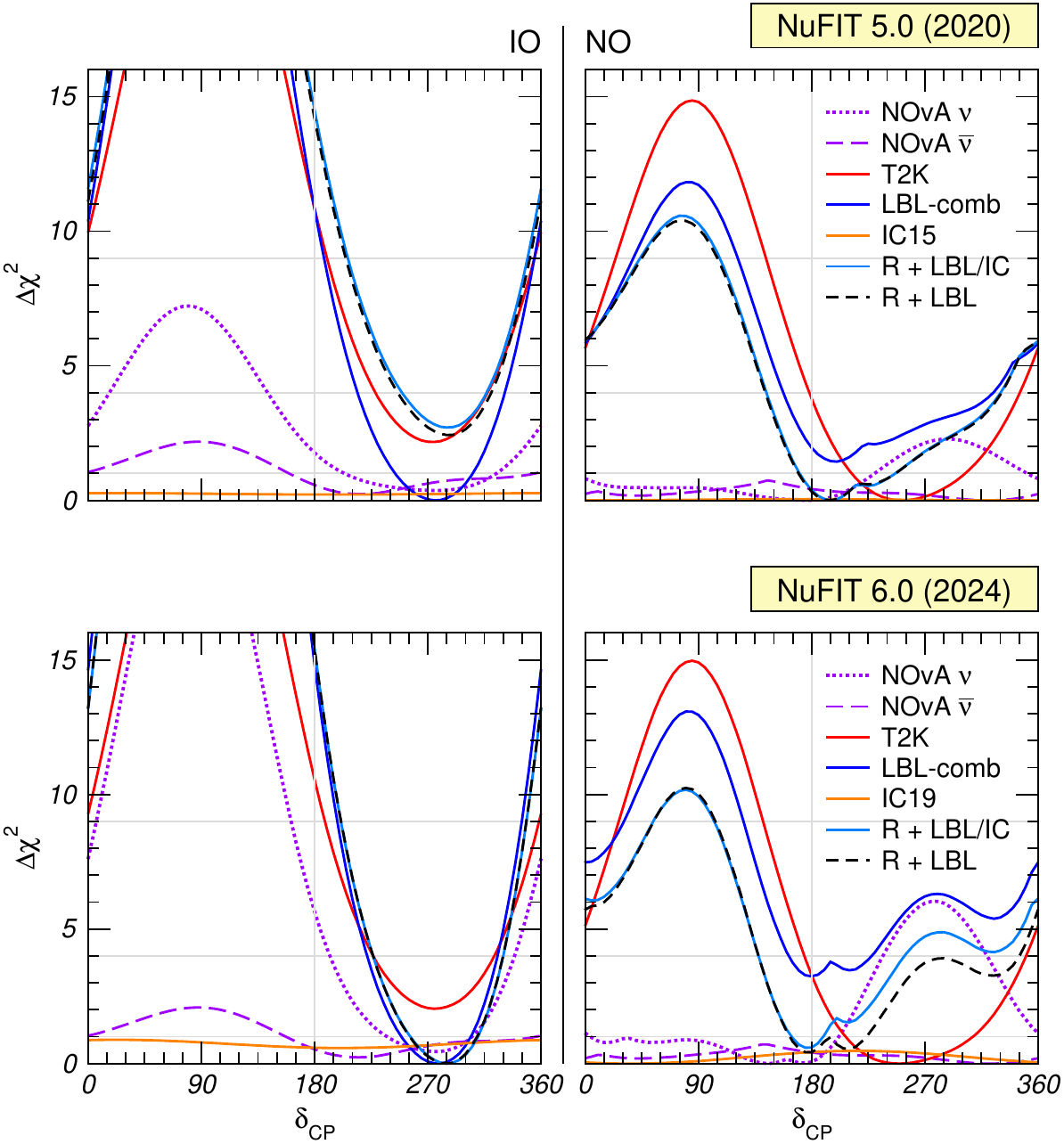}
  \caption{$\Delta\chi^2$ profiles as a function of $\dCP$ for
    different data sets and combinations as labeled in the figure.  In
    the curves where the reactors $R$ are not included in the
    combination we have fixed $\sin^2\theta_{13} = 0.0222$ as well as
    the solar parameters and minimized with respect to $\theta_{23}$
    and $|\Dmq_{3\ell}|$.  When the reactors are included
    $\theta_{13}$ is also marginalized.  Left (right) panels are for
    IO (NO) and $\Delta\chi^2$ is shown with respect to the global
    best-fit point for each curve.  Upper panels are for the NuFIT~5.0
    data set, whereas lower panels correspond to the current update.}
  \label{fig:compare-dcp24}
\end{figure}

This is further illustrated in Fig.~\ref{fig:compare-dcp24}, which
shows the $\Delta\chi^2$ profiles as a function of $\dCP$ for the LBL
experiments T2K and NOvA and their combination (we also add the
information from MBL reactors and from the IC atmospheric samples
which we independently analyzed in NuFIT~5.0 and NuFIT~6.0, and will
discuss in Sec.~\ref{sec:reacatm}).  Comparing the curves of T2K and
NOvA in the upper and lower panels, we see that the main difference is
in the NOvA neutrino results.  Just by themselves, they disfavor NO
and $\dCP\simeq 3\pi/2$ by about 6 units in $\chi^2$ (versus 3 units
in NuFIT~5.0), whereas comparing the corresponding curves in the left
panels we see that for IO the consistent preference of T2K and NOvA
for $\dCP\simeq 3\pi/2$ is now statistically more significant.
Altogether, this drives the preference of the LBL combination for IO,
with $\Delta\chi^2(\text{NO}-\text{IO}) \approx 3.2$ versus 1.5 in
NuFIT~5.0.

\begin{figure}\centering
  \includegraphics[width=\textwidth]{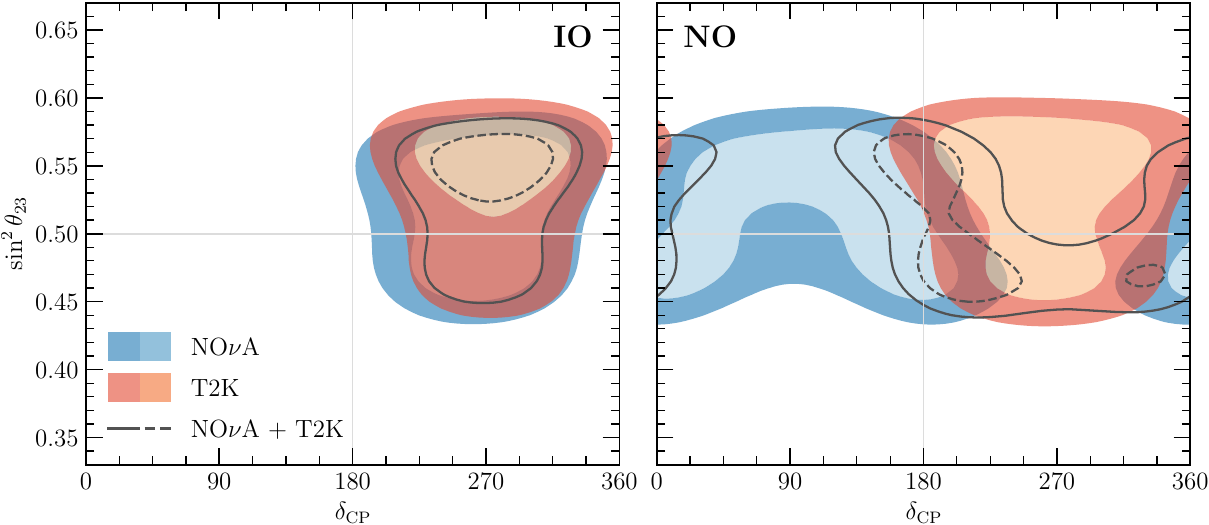}
  \caption{$1\sigma$ and $2\sigma$ allowed regions (2 dof) for T2K
    (red shading), NOvA (blue shading) and their combination (black
    curves).  Contours are defined with respect to the local minimum
    for IO (left) or NO (right).  We fix $\sin^2\theta_{13}=0.0222$,
    $\sin^2\theta_{12}=0.31$, $\Dmq_{21}=7.5\times 10^{-5}~\eVq$ and
    minimize with respect to $|\Dmq_{3\ell}|$.}
  \label{fig:sq23-dCP24}
\end{figure}

\begin{figure}\centering
  \includegraphics[width=0.9\textwidth]{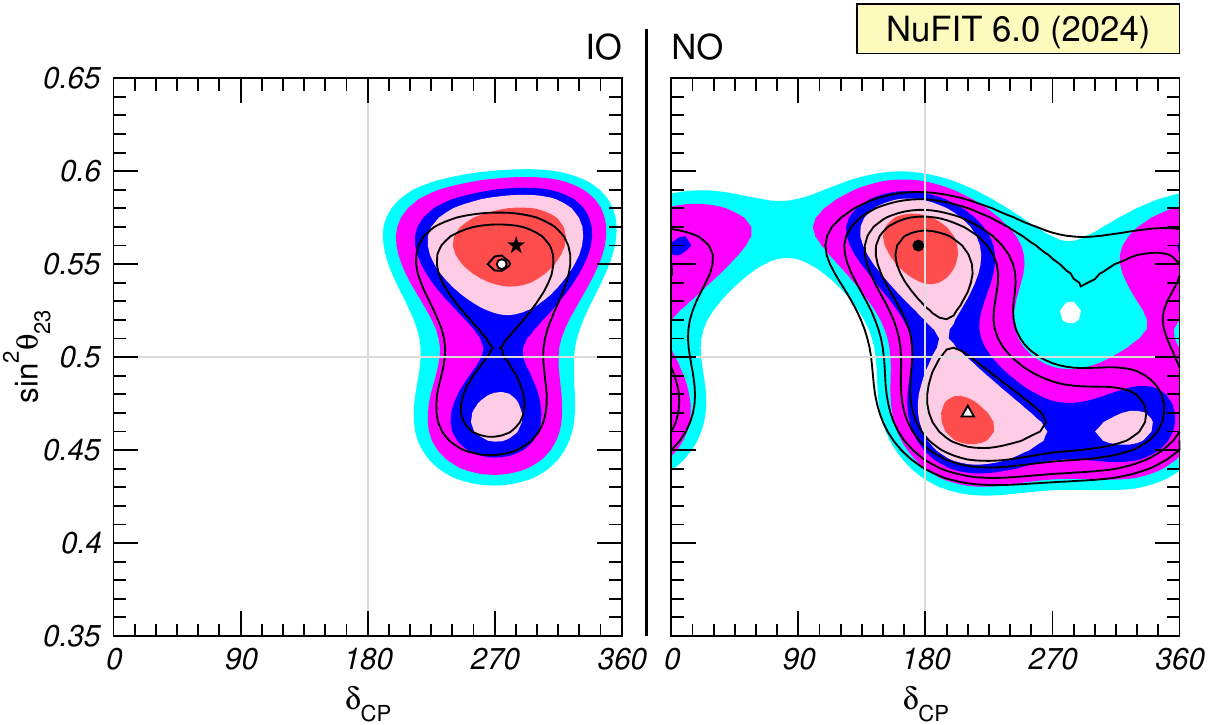}
  \caption{Two-dimensional projection of the allowed six-dimensional
    region from global data in the plane of ($\dCP,\sin^2\theta_{23}$)
    for IO (left) and NO (right) after minimization with respect to
    the undisplayed parameters.  Regions for both orderings are
    defined with respect to the global best-fit point.  The different
    contours correspond to $1\sigma$, 90\%, $2\sigma$, 99\%, $3\sigma$
    CL (2 dof).  Colored regions (black contours) correspond to the
    variant with IC19 and without SK-atm (with IC24 and with SK-atm).}
  \label{fig:region-cp23}
\end{figure}

The two-dimensional regions for T2K and NOvA in the ($\dCP$,
$\sin^2\theta_{23}$) plane for fixed $\theta_{13}$ are shown in
Figure~\ref{fig:sq23-dCP24}.  The better consistency for IO is
apparent.  For NO we find that, unlike in NuFIT~5.0, the $1\sigma$
regions do not overlap.  The 2-dimensional projections in the ($\dCP$,
$\sin^2\theta_{23}$) plane from the global analysis of all data are
shown in Fig.~\ref{fig:region-cp23}, which resemble to a large extent
the features from the combination among T2K and NOvA discussed above.
We observe, in particular, non-trivial correlations between these two
parameters and the MO.  For IO, the preference for $\dCP \simeq
270^\circ$ is highly significant, whereas for NO a more complicated
structure in the ($\dCP$, $\sin^2\theta_{23}$) plane, with several
local minima, emerges.  The octant degeneracy for $\theta_{23}$ is
present with $\Delta\chi^2< 4$ for both mass orderings and both data
variants, showing local minima around $\sin^2\theta_{23} \approx 0.56$
and 0.47.

An obvious question to address is whether T2K and NOvA are in tension
with each other at a worrisome level.  Consistency among different
data sets can be quantified with the parameter goodness-of-fit
(PG)~\cite{Maltoni:2003cu}.  For a number $N$ of different data sets
$i$, each depending on $n_i$ model parameters, and globally depending
on $n_\text{glob}$ parameters, it can be shown that the test statistic
\begin{equation}
  \chi^2_\text{PG} \equiv \chi^2_\text{min,glob} - \sum_i^N
  \chi^2_{\text{min}, i} = \min\bigg[ \sum_i^N\chi^2_i \bigg]
  - \sum_i \chi^2_{\text{min}, i} \,,
\end{equation}
follows a $\chi^2$ distribution with $n \equiv \sum_i n_i -
n_\text{glob}$ degrees of freedom~\cite{Maltoni:2003cu}.

Applying this test to the full NOvA and T2K samples (including both
appearance and disappearance data for neutrinos and antineutrinos) we
obtain the values in Table~\ref{tab:PG24}.  We carry out the analysis
separately for each mass ordering, in all cases fixing $\Dmq_{21}$ and
$\theta_{12}$ to their best fit.  In the results reported in the upper
part of the table $\theta_{13}$ is varied in the minimization, so
$n_\text{T2K} = n_\text{NOvA} = n_\text{glob=T2K+NOvA} = 4$
(\textit{i.e.}, $\Dmq_{3\ell}$, $\theta_{23}$, $\dCP$, and
$\theta_{13}$).  In the lower part $\theta_{13}$ is kept fixed to its
best fit so $n_\text{T2K} = n_\text{NOvA} = n_\text{glob=T2K+NOvA} =
3$.  From the table we read that, as expected, agreement is better in
IO, where irrespective on $\theta_{13}$ the samples are compatible at
the $0.5\sigma$ level or better.  In NO, compatibility arises at
$1.7\sigma$ ($2.0\sigma$) for free (fixed) $\theta_{13}$.  This is to
be compared with the NuFIT~5.0 results of $1.4\sigma$ ($1.7\sigma$)
respectively.  We conclude that the tension between T2K and NOvA in NO
has slightly strengthened with the new results, reaching at most the
$2\sigma$ level.

\begin{table}\centering
  \catcode`?=\active\def?{\hphantom{0}}
  \catcode`!=\active\def!{\hphantom{.}}
  \begin{tabular}{l|ccc|ccc}
  \hline\hline
  Data sets &  \multicolumn{3}{c|}{Normal Ordering}
  &  \multicolumn{3}{c}{Inverted Ordering} \\
  & $\chi^2_\text{PG} / n$ & $p$-value & $\#\sigma$
  & $\chi^2_\text{PG} / n$ & $p$-value & $\#\sigma$ \\
  \hline
  T2K vs NOvA                  & ?7.9/4? & 0.093 & 1.7? & ?2.3/4? & 0.67 & 0.42 \\
  T2K vs React                 & 0.23/2? & 0.89? & 0.14 & ?1.7/2? & 0.43 & 0.79 \\
  NOvA vs React                & ?1.1/2? & 0.58? & 0.56 & ?4.3/2? & 0.12 & 1.6? \\
  T2K vs NOvA vs React         & ?8.6/6? & 0.20? & 1.3? & ?6.0/6? & 0.42 & 0.80 \\
  (T2K \& NOvA) vs React       & 0.76/2? & 0.68? & 0.41 & ?3.4/2? & 0.18 & 1.3? \\
  T2K vs IC19                  & ?2.7/4? & 0.61? & 0.51 & ?1.2/4? & 0.88 & 0.15 \\
  NOvA vs IC19                 & ?3.3/4? & 0.51? & 0.66 & ?2.3/4? & 0.68 & 0.41 \\
  Reac vs IC19                 & ?2.1/2? & 0.35? & 0.93 & 0.88/2? & 0.64 & 0.84 \\
  NOvA vs T2K vs IC19          & !?11/8? & 0.20? & 1.3? & ?4.3/8? & 0.83 & 0.21 \\
  NOvA vs T2K vs React vs IC19 & 11.5/10 & 0.33? & 0.96 & ?7.2/10 & 0.71 & 0.38 \\
  \hline
  T2K vs NOvA                  & ?8.0/3 & 0.045 & 2.0? & 1.8/3 & 0.61? & 0.50? \\
  T2K vs NOvA vs React         & ?8.3/4 & 0.081 & 1.7? & 4.1/4 & 0.39? & 0.85? \\
  (T2K \& NOvA) vs React       & 0.25/1 & 0.62? & 0.50 & 2.0/1 & 0.16? & 1.4?? \\
  T2K vs IC19                  & 0.72/3 & 0.86? & 0.16 & 0.2/3 & 0.98? & 0.028 \\
  NOvA vs IC19                 & ?1.5/3 & 0.68? & 0.41 & 1.0/3 & 0.80? & 0.25? \\
  NOvA vs T2K vs IC19          & ?9.3/6 & 0.16? & 1.4? & 2.4/6 & 0.88? & 0.15? \\
  NOvA vs T2K vs React vs IC19 & ?9.4/7 & 0.22? & 1.2? & 4.5/7 & 0.72? & 0.36? \\
  NOvA vs T2K vs IC24          & ?9.5/6 & 0.15? & 1.4? & 4.4/6 & 0.62? & 0.49? \\
  NOvA vs T2K vs React vs IC24 & !?10/7 & 0.19? & 1.3? & 8.2/7 & 0.27? & 1.1?? \\
  \hline\hline
  \end{tabular}
  \caption{Consistency test among different data sets, shown in the
    first column, assuming either normal or inverted ordering.
    ``React'' includes Daya-Bay, RENO and Double-Chooz.  In the
    analyses above the horizontal line, $\theta_{13}$ is a free
    parameter, whereas below the line we have fixed $\sin^2\theta_{13}
    = 0.0222$.  See text for more details.}
    \label{tab:PG24}
\end{table}

\subsection{Effects from \texorpdfstring{$\nu_\mu/\bar\nu_\mu$}{numu/antinumu} versus \texorpdfstring{$\bar\nu_e$}{antinue} disappearance}
\label{sec:reacatm}

\begin{figure}\centering
  \includegraphics[width=0.9\textwidth]{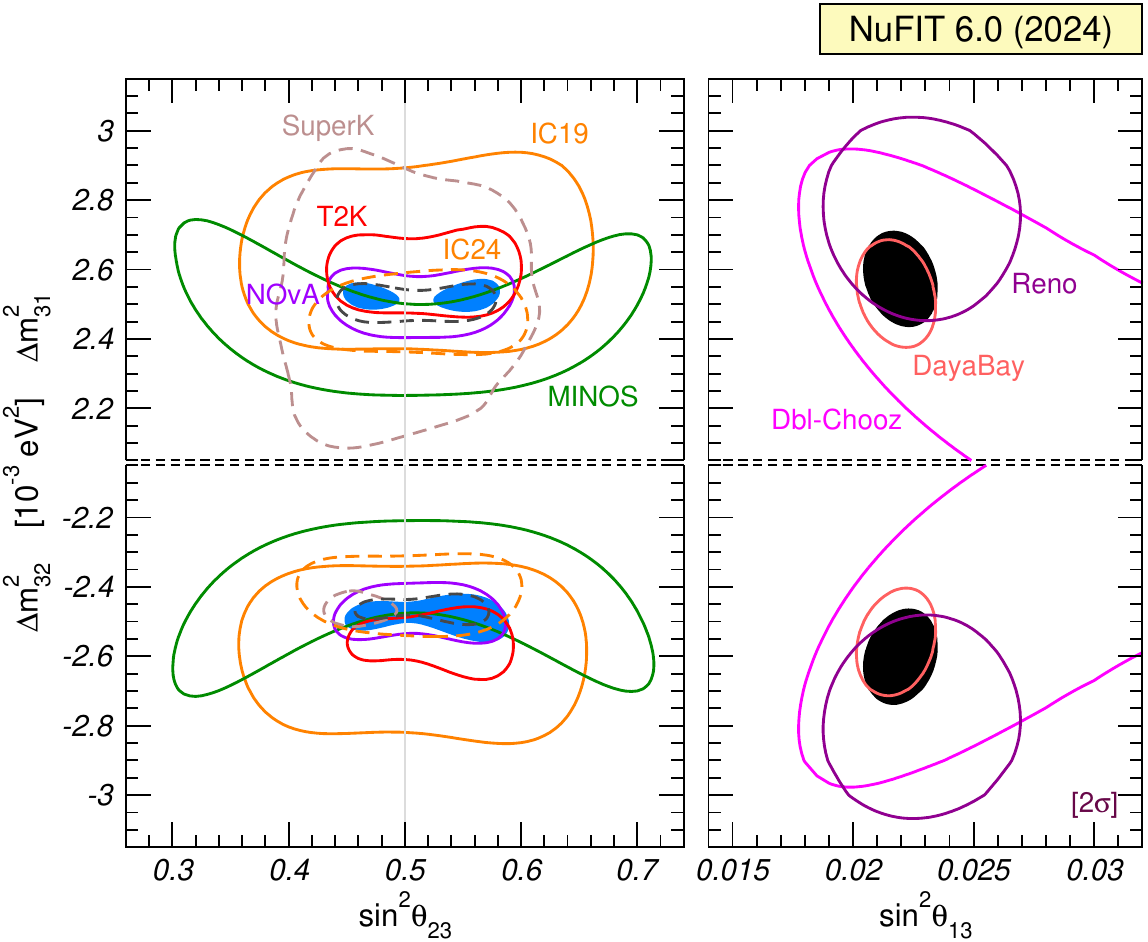}
  \caption{Confidence regions at 95.45\% CL (2~dof) in the plane of
    $\sin^2\theta_{23}$ ($\sin^2\theta_{13}$) and $\Dmq_{3\ell})$ in
    the left (right) panels.  For the left panels we use both
    appearance and disappearance data from MINOS (green), NOvA
    (purple) and T2K (red), as well as atmospheric data from IC
    (orange) and Super-Kamiokande (light-brown); the colored region
    corresponds to the combination of these accelerator data with
    IC19, whereas the black-dashed contour corresponds to the
    combination with IC24 and Super-Kamiokande.  A prior on
    $\theta_{13}$ is included to account for the reactor constraint.
    The right panels show regions using data from Daya-Bay (pink),
    Double-Chooz (magenta), RENO (violet), and their combination
    (black regions).  In all panels solar, KamLAND and SNO+ data are
    included to constrain $\Dmq_{21}$ and $\theta_{12}$.  Contours are
    defined with respect to the global minimum of the two orderings
    for each data set.}
  \label{fig:regions-dis}
\end{figure}

Figure~\ref{fig:regions-dis} shows the combined determination of the
parameters $\sin^2\theta_{23}$, $\sin^2\theta_{13}$, and
$\Dmq_{3\ell}$ by the interplay of different data samples, namely
$\nu_\mu/\bar\nu_\mu$ disappearance from long-baseline accelerator and
atmospheric neutrino data (left panel) and $\bar\nu_e$ disappearance
from medium baseline reactor experiments (right panel).  We observe
significant synergy from the combination of different data sets
(global regions are clearly smaller than individual ones) as well as
appealing consistency.  This is also reflected by the many
compatibility tests reported in Table~\ref{tab:PG24}, which all show
very good consistency typically well below $2\sigma$.

\begin{figure}\centering
  \includegraphics[width=0.49\textwidth]{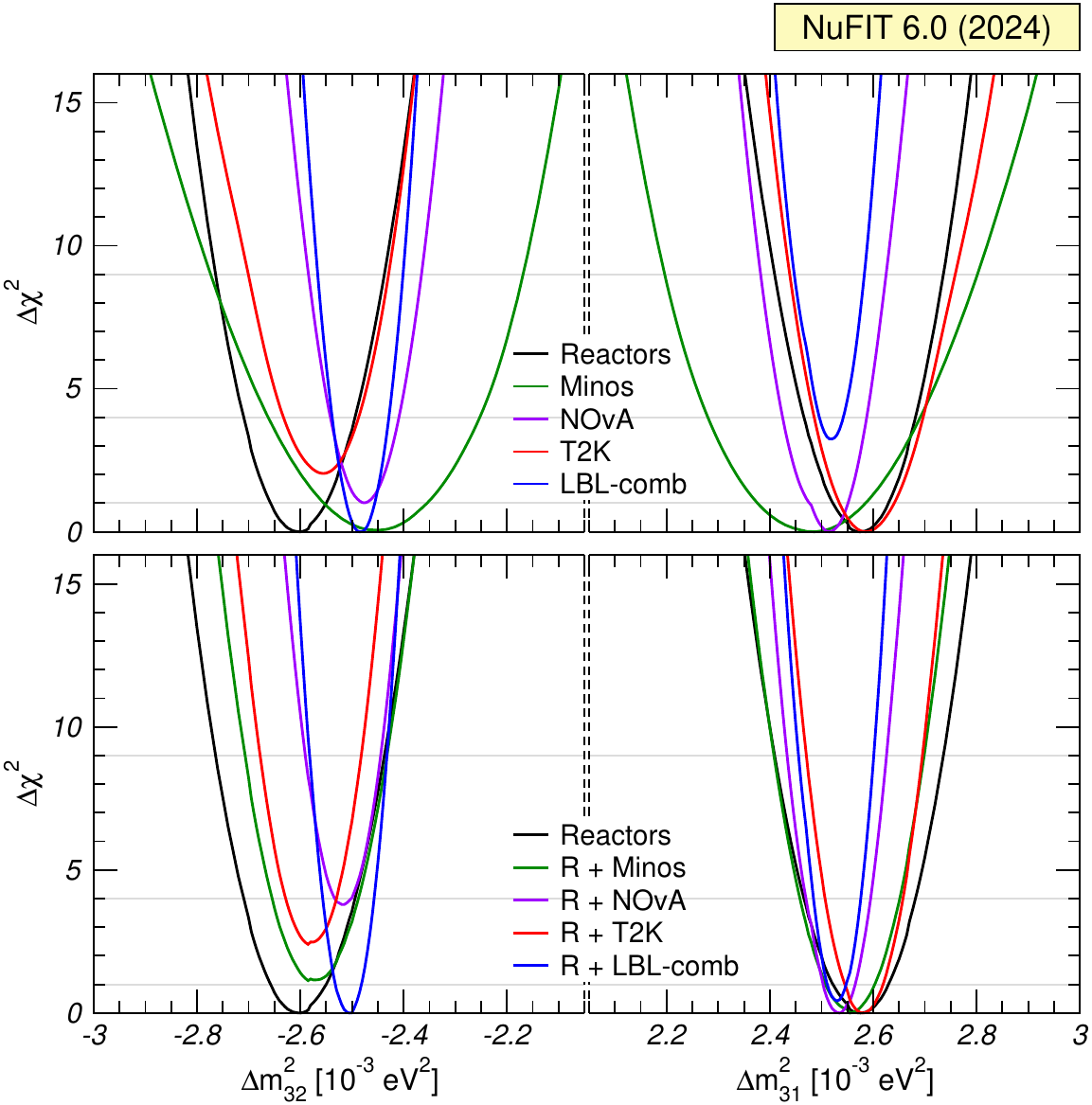}\hfill
  \includegraphics[width=0.49\textwidth]{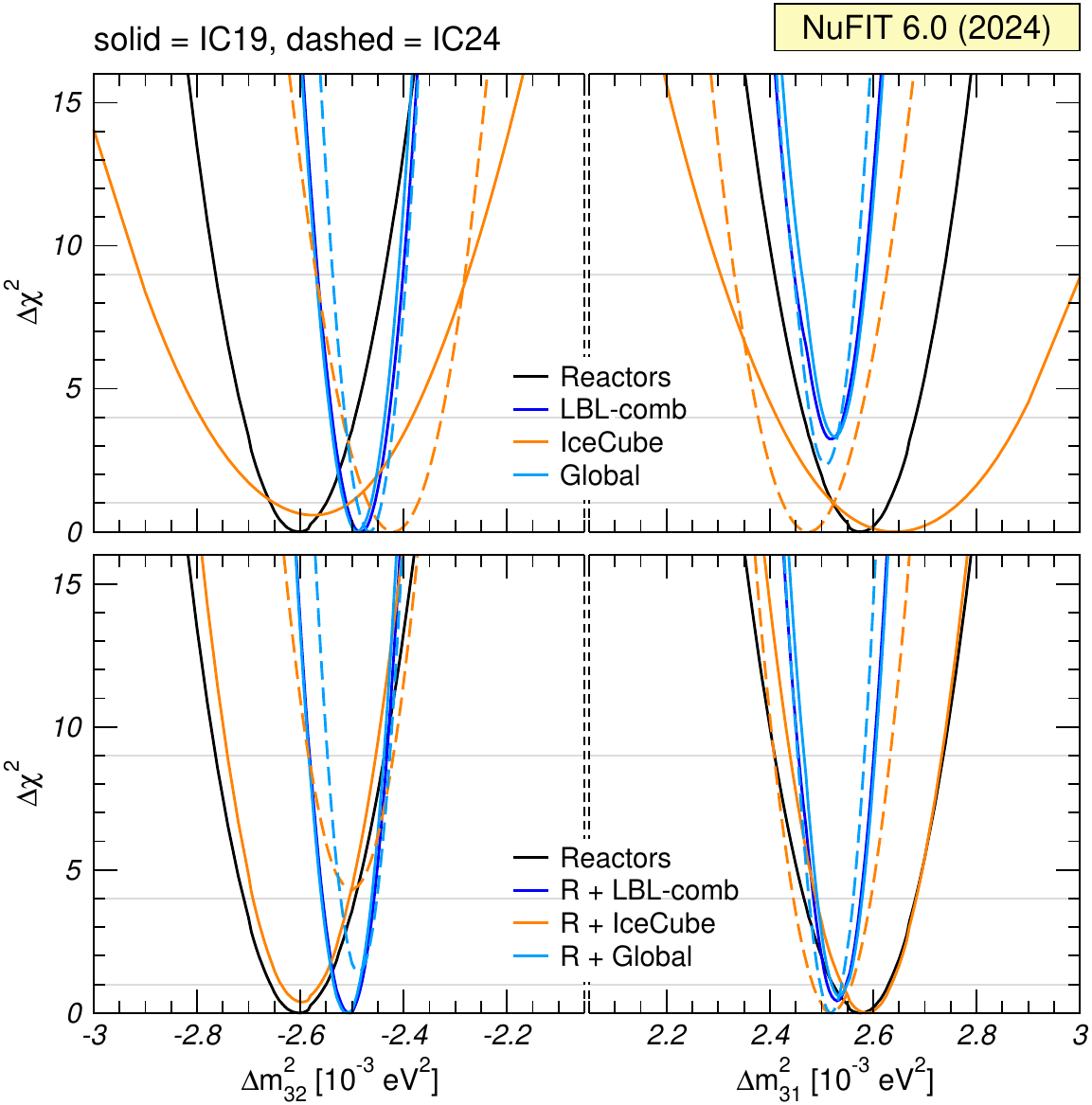}
  \caption{$\Delta\chi^2$ profiles as a function of $\Dmq_{3\ell}$ for
    different data sets and combinations as labeled in the figures.
    In the curves where the reactors $R$ are not included in the
    combination we have fixed $\sin^2\theta_{13}=0.0222$ as well as
    the solar parameters and minimized with respect to $\theta_{23}$
    and $\dCP$.  When the reactors are included $\theta_{13}$ is also
    marginalized.  $\Delta\chi^2$ is shown with respect to the global
    best-fit point (IO or NO) for each curve.  The left set of panels
    visualizes the reactor/LBL combination, whereas in the right set
    of panels we are illustrating the impact of the IC19 or IC24 data
    sets.}
  \label{fig:chisq-dma}
\end{figure}

As in previous analyses, the combination of $\nu_\mu$ and
$\bar\nu_\mu$ disappearance data (dominated by LBL accelerator
experiments) with $\bar\nu_e$ disappearance from reactors provides
complementary information, which is especially relevant for the MO
discrimination~\cite{Nunokawa:2005nx, Minakata:2006gq}.  In short, the
effective mass-squared difference relevant for each survival
probability depends on the mass ordering, so a global combination
allows in principle to determine it.  This effect can be seen in
1-dimensional $\chi^2$ projections on the parameter $\Dmq_{3\ell}$, as
shown in Fig.~\ref{fig:chisq-dma}.

In the upper panels we show the profiles for various individual data
sets.  We observe that, individually, the considered data are either
insensitive to the MO ($\Delta\chi^2 < 1$) or show a slight preference
for NO.  The only exception is the LBL/IC combination, which prefers
IO due to the T2K/NOvA tension discussed above.  However, by comparing
the LBL and reactor results, we observe that the determination of
$|\Dmq_{3\ell}|$ is in better agreement for NO than for IO.  Hence,
combining reactor and LBL data (bottom-left panels) increases the
value of $\Delta\chi^2_\text{IO,NO} \equiv \chi^2_\text{min,IO} -
\chi^2_\text{min,NO}$ in favor of NO.  As a result, the preference for
IO from the T2K/NOvA combination (dominated by appearance data) is
nearly exactly compensated by the effects of disappearance data in the
LBL/reactor combination, leading to a global result of
$\Delta\chi^2_\text{IO,NO} = -0.6$.

The preference for NO from the accelerator/reactor combination is also
visible in the PG tests in Table~\ref{tab:PG24}, by considering the
consistency among the combined T2K \& NOvA sample and reactors.  For
both $\theta_{13}$ free and fixed, there is slightly better
compatibility for NO than IO (although even for IO consistency is very
good, $1.3\sigma$ or $1.4\sigma$).  This preference is only visible
when combining T2K and NOvA \emph{before} testing the consistency
(\textit{i.e.}, the rows labeled <<(T2K \& NOvA) vs React>>).
Otherwise, if they are kept separate (\textit{i.e.}, the rows labeled
<<T2K vs NOvA vs React>>) the opposite trend from the combination
among T2K and NOvA appearance samples compensates for the tendency
from disappearance data.

In the right half of Fig.~\ref{fig:chisq-dma}, we show the impact of
the different IceCube data samples.  It is clear from the figure that
the IC19 3-year data sample~\cite{IceCube:2019dyb, IceCube:2019dqi}
plays very little role in the $\nu_\mu/\nu_e$ disappearance
complementarity due to its relatively weak constraint on
$|\Dmq_{3\ell}|$.  However, when combining the IC24 $\chi^2$ table
corresponding to 9.3 years of data~\cite{IceCube:2024xjj, IC:data2024}
with reactor data, this complementarity \cite{Blennow:2013vta} already
provides a preference for NO, with $\Delta\chi^2_\text{IO,NO} \approx
4.5$.  The result is entirely driven by the $|\Dmq_{3\ell}|$
determination, as the $\chi^2$ tables provided by the collaboration
contain no information on the IceCube MO sensitivity (they only
provide relative $\Delta\chi^2$ values with respect to the best fit
\emph{in each ordering}).  Let us remark, however, that the
$\Delta\chi^2_\text{IO,NO} = 4.5$ contribution from combining IC24 and
reactors does not simply add up to the value
$\Delta\chi^2_\text{IO,NO} = -0.6$ from combining LBL and reactors.
Instead, combining LBL and IC24 leads to a shift in $|\Dmq_{3\ell}|$
that, when adding reactor data, leads to $\Delta\chi^2_\text{IO,NO}
\approx 1.5$.

Different to the IC24 data table, the latest Super-Kamiokande
atmospheric data~\cite{SKatm:data2024} alone shows a preference for NO
with $\Delta\chi^2_\text{IO,NO} \approx 5.7$.  We note, however, that
this result seems to emerge from a large statistical fluctuation.
Indeed, the probability of obtaining the data is relatively low for
both mass orderings, and considering the distribution of the relevant
test statistic, the SK collaboration determines a preference for NO
over IO at the 92.3\%~CL~\cite{Super-Kamiokande:2023ahc}.  When
combining the IC24 and SK atmospheric neutrino $\chi^2$ tables with
our global fit of the remaining data, we find an overall preference
for NO with $\Delta\chi^2_\text{IO,NO} \approx 6.1$, see
Sec.~\ref{sec:global24}.

\subsection{Sensitivity to the neutrino mass ordering}
\label{sec:MO}

Given the different trends among several determinations of the mass
ordering, we now study in more detail the sensitivity of current
global data to it.  To do so, we follow the methodology in
Ref.~\cite{Blennow:2013oma}.  As customary, a useful test statistic
for this purpose is the $\chi^2$ difference among the best-fit points
for the two orderings.  Following Ref.~\cite{Blennow:2013oma}, we
denote it in this Section as $T$,
\begin{equation}
  T \equiv \Delta\chi^2_\text{IO,NO}
  \equiv \chi^2_\text{min,IO} - \chi^2_\text{min,NO} \,.
\end{equation}
Hence, positive values of $T$ favor NO, and negative values favor IO.
As shown in refs.~\cite{Blennow:2013oma, Qian:2012zn}, under certain
conditions $T$ will follow a Gaussian distribution with mean $\pm T_0$
and standard deviation $2\sqrt{T_0}$, where $T_0$ is obtained as
follows.  If $p_i(o, \theta)$ is the expected number of events in bin
$i$ (where $o \in \{\text{NO}, \text{IO}\}$ is the mass ordering and
$\theta$ are the remaining oscillation parameters), $d_i$ is the
observed number of events in that bin, and the global $\chi^2$ is
given by $\chi^2 = \chi^2\big[ p_i(o, \theta);\, d_i \big]$, then
\begin{equation}
  \label{eq:T0def}
  \begin{split}
    T_0^\text{NO} \equiv \text{min}_{\theta} \Big\{
    \chi^2\big[ p_i(\text{IO},\theta);\, p_i(\text{NO},\theta^\text{true}) \big]
    \Big\} \,,
    \\
    T_0^\text{IO} \equiv \text{min}_{\theta} \Big\{
    \chi^2\big[ p_i(\text{NO},\theta);\, p_i(\text{IO},\theta^\text{true}) \big]
    \Big\} \,.
  \end{split}
\end{equation}
That is, $T_0$ is determined by replacing the data by the prediction
for the opposite mass ordering, given some assumed true values of
oscillation parameters.  Defined this way, $T_0^o$ is always positive.
Hence, for true NO (IO) the expected value of $T$ is $+T_0^\text{NO}$
($-T_0^\text{IO}$).  The conditions under which the aforementioned
Gaussian approximation holds are discussed in detail in the Appendix
of Ref.~\cite{Blennow:2013oma}, and they are similar to the conditions
for Wilks' theorem.  In the following, we will study the results of
the global fit under the assumption that $T$ is indeed
Gaussian-distributed.  In addition, the values of $T_0^o$ defined in
Eq.~\eqref{eq:T0def} depend on the (unknown) true values of the
oscillation parameters $\theta^\text{true}$.  In the following, we
will take them to be the best-fit points of our analysis as given in
Table~\ref{tab:bfranges24}; we comment on this assumption in
Appendix~\ref{sec:app-T}.

\begin{figure}\centering
  \includegraphics[width=0.9\textwidth]{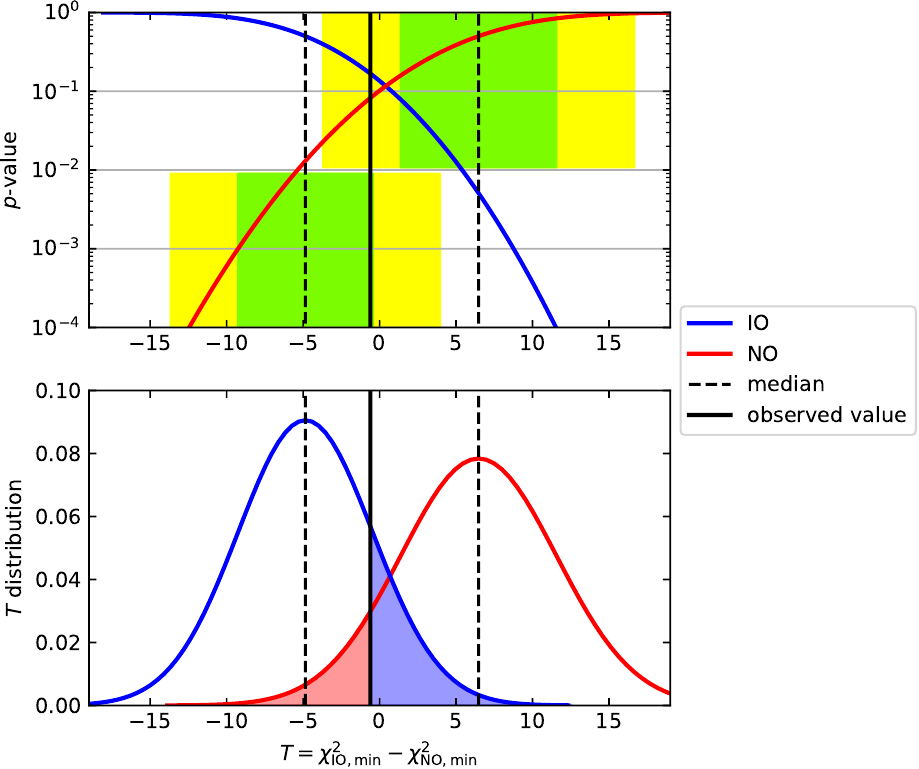}
  \caption{$p$-values (top) and distributions (bottom) for the test
    statistic $T = \chi^2_\text{IO,min} - \chi^2_\text{NO,min}$
    corresponding to the <<IC19 w/o SK-atm>> analysis, assuming true
    NO (red) or true IO (blue).  The observed value $T_\text{obs} =
    -0.6$ is shown by the solid vertical black line.  The
    corresponding median values are shown by the dashed vertical
    lines.  The green and yellow bands in the top panel ---vertically
    displaced to avoid graphical overlap--- correspond to the
    $1\sigma$ and $2\sigma$ intervals for $T$ assuming NO (upwards
    displaced bands) and IO (downwards displaced bands).}
  \label{fig:T-test}
\end{figure}

This analysis cannot be carried out for the atmospheric neutrino data
samples IC24 and Super-Kamiokande, which are included in our global
comprehensive fit as numerical $\chi^2$ tables.
Therefore, in what follows we only
consider the data sample denoted by <<IC19 w/o SK-atm>>.\footnote{This
is unfortunate, as the IC24 and SK samples provide relevant
sensitivity to the MO (see above).} For the global data combination,
we find
\begin{equation}
  \label{eq:T0}
  T_0^\text{NO} = 6.47 \,,
  \qquad
  T_0^\text{IO} = 4.85 \,.
\end{equation}
The corresponding normal distributions for $T$ are shown in the bottom
panel of Fig.~\ref{fig:T-test}.  For an observed value $T_\text{obs}$,
we compute the $p$-value for a given mass ordering as usual: the
$p$-value for IO (NO) is the probability to obtain a value of $T$
larger (smaller) than $T_\text{obs}$ if the true ordering is IO (NO).
Given the observed value for the global fit, $T_\text{obs} = -0.6$,
the corresponding $p$-values are indicated by shaded areas in the
lower panel of Fig.~\ref{fig:T-test} and can be read off from the
intersection of the red and blue curves with the $T_\text{obs}$ value
in the upper panel:
\begin{equation}
  \begin{aligned}
    &\text{NO:}
    & p_\text{NO} &= \hphantom{0}8.2\% \,,
    && 91.8\%~\text{CL} \,,
    && 1.7\sigma \,,
    \\
    &\text{IO:}
    & p_\text{IO} &= 16.7\% \,,
    && 83.3\%~\text{CL} \,,
    && 1.4\sigma \,.
  \end{aligned}
\end{equation}
As expected for $|T_\text{obs}| < 1$, the $p$-values for both
orderings are similar and we cannot significantly favor one over the
other.  Indeed, none of the orderings can be rejected at relevant
significance; we obtain $p$-values below $2\sigma$ for both orderings,
with a slightly smaller $p$-value for NO due to the negative value of
$T_\text{obs}$.

Given the distribution for $T$, we can estimate the sensitivity of the
considered data set.  In particular, the median sensitivity is
obtained by assuming that $T_\text{obs}$ is given by the mean value of
$T$ for a given mass ordering.  Hence, the median p-value is given by
the intersection of the dashed lines with the corresponding red or
blue curves in the top panel of Fig.~\ref{fig:T-test},
\begin{equation}
  \begin{aligned}
    &\text{NO:}
    & p_\text{NO}^\text{med} &= 1.3\% \,,
    && \hphantom{0}98.7\%~\text{CL} \,,
    && 2.5\sigma \,,
    \\
    &\text{IO:}
    & p_\text{IO}^\text{med} &= 0.51\% \,,
    && 99.49\%~\text{CL} \,,
    && 2.8\sigma \,.
  \end{aligned}
\end{equation}
We conclude that current data has a nominal sensitivity above
$2.5\sigma$ to the mass ordering.  The weak rejection we obtain for
both orderings is a result of the opposite trends in the data
discussed in previous subsections, resulting in an observed value for
$T_\text{obs}$ right in between the peaks of the distributions.  A
natural question is how unlikely this result is.  To assess it, we
show in the top panel of Fig.~\ref{fig:T-test} the intervals where
$T_\text{obs}$ is expected to lie with probability of 68.27\% (green)
and 95.45\% (yellow) for the two mass orderings.  We see that the
obtained value $T_\text{obs} = -0.6$ is not particularly unlikely for
both orderings, being located within the $1\sigma$ ($2\sigma$) ranges
for IO (NO).

\section{Updates in the ``12'' sector}
\label{sec:solar}

\begin{figure}\centering
  \includegraphics[width=0.9\textwidth]{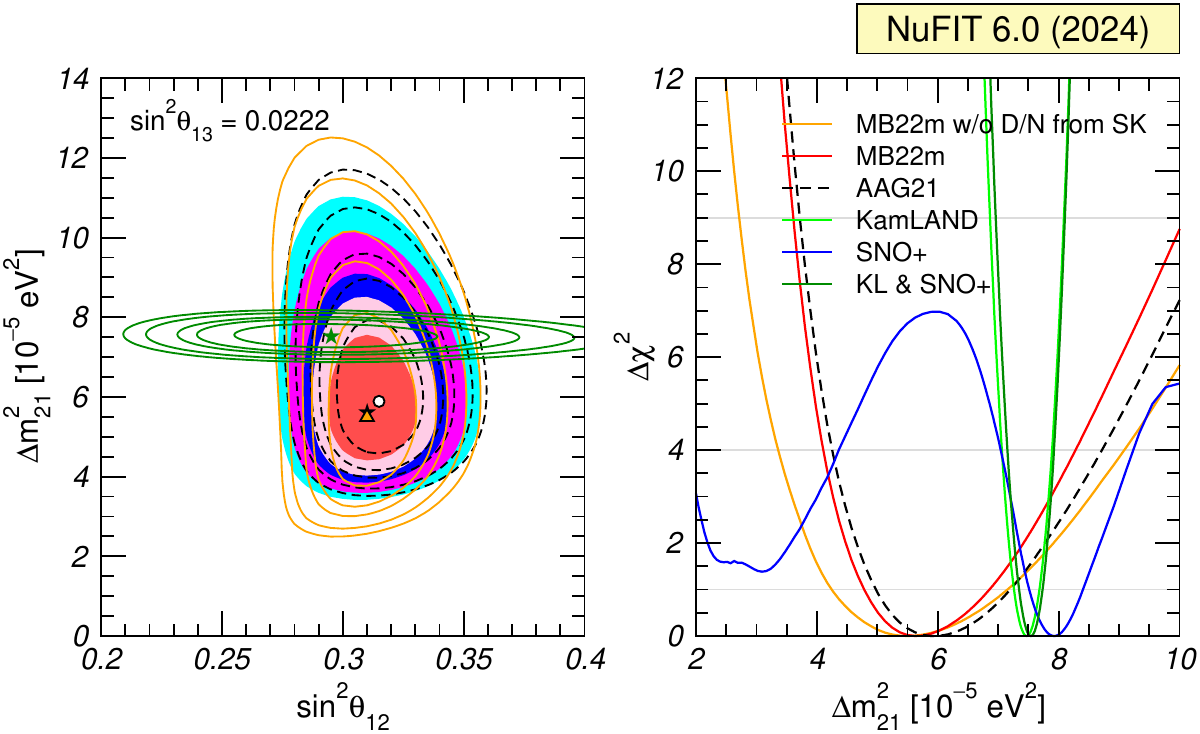}
  \caption{Left: Allowed parameter regions (at $1\sigma$, 90\%,
    $2\sigma$, 99\%, and $3\sigma$ CL for 2 dof) from the combined
    analysis of solar data for MB22-met model (full regions with best
    fit marked by black star) and AAG21 model (dashed void contours
    with best fit marked by a white dot), and for the analysis of the
    combination of KamLAND and SNO+ data (solid green contours with
    best fit marked by a green star) for fixed
    $\sin^2{\theta_{13}}=0.0222$.  For comparison we also show as
    orange contours the results obtained with the MB22-met model
    without including the results of the day-night variation in SK.
    Right: $\Delta\chi^2$ dependence on $\Dmq_{21}$ for the same four
    analyses after marginalizing over $\theta_{12}$.  In addition we
    show separately the results from KamLAND and SNO+.}
  \label{fig:sun-tension24}
\end{figure}

The analyses of the solar experiments and of reactor experiments at
$\mathcal{O}(100~\text{km})$ distance (which we refer to as LBL
reactor experiments) give the dominant contribution to the
determination of $\Dmq_{21}$ and $\theta_{12}$.  We show in
Fig.~\ref{fig:sun-tension24} the present determination of these
parameters from the global solar analysis in comparison with that of
LBL reactor data.

In the solar neutrino sector, new data included since NuFIT~5.0 are
the full day-night spectrum from the phase-IV of
Super-Kamiokande~\cite{Cravens:2008aa}, and the final spectra from
Borexino phases-II~\cite{Borexino:2017rsf} and
III~\cite{BOREXINO:2022abl}.  As for the predictions required for the
solar neutrino analysis, the main update is that we have employed the
new generation of Standard Solar Models~\cite{B23Fluxes}.  In brief,
for the last two decades solar modeling has suffered from the
so-called solar composition problem, associated with the choice of the
input for heavy element abundances.  They were either taken from the
older results from Ref.~\cite{Grevesse1998} (GS98), which implied a
higher metallicity and predicted solar properties in good agreement
with helioseismology observations, or the newer abundances (obtained
with more modern methodology and techniques) summarized in
Ref.~\cite{Asplund2009} (AGSS09), which implied a lower metallicity
and did not agree with helioseismology.  Consequently, two different
sets of Standard Solar Models were built, each based on the
corresponding set of solar abundances~\cite{Serenelli:2009yc,
  Serenelli:2011py, Vinyoles:2016djt}.  On this front, an update of
the AGSS09 results was presented by the same group
(AAG21)~\cite{Asplund2021}, leading only to a slight increase of the
solar metallicity.  On top of that, a new set of results
(MB22)~\cite{Magg:2022rxb} ---~based on similar methodologies and
techniques but with different atomic input data for the critical
oxygen lines, among other differences~--- led to a substantial change
in solar element abundances with respect to AGSS09, more in agreement
with those from GS98.  Therefore, the models built following MB22
provide a good description of helioseismology results.

In Fig.~\ref{fig:sun-tension24} we show the present determination of
$\Dmq_{21}$ and $\theta_{12}$ from the global solar analysis performed
with the two extreme versions of the Standard Solar Model, namely the
one based on the AAG21 abundances and the one based on the MB22-met
abundances.  From the figure, we see that the determination of the
parameters is robust under the changes in the modeling, though the
allowed ranges ---~particularly at higher CL~--- is different for the
two models.  In this respect, it is important to point out that the
latest Standard-Solar-Model-independent determination of the solar
fluxes performed in Ref.~\cite{Gonzalez-Garcia:2023kva} shows better
agreement with the predictions of the MB22 models.  For this reason,
we adopt the MB22 as the reference model employed for the results
reported in NuFIT~6.0.

In what respects the relevant data from reactor experiments, we have
updated the antineutrino fluxes used for the predictions to the latest
Daya-Bay measurements~\cite{DayaBay:2021dqj}.  Furthermore, we have
included in the fit our analysis of the first results reported by the
SNO+ collaboration, which combines the 114 ton-yr of data gathered
during the partial-fill phase reported in Ref.~\cite{SNO:2024wzq} and
the first 286 ton-yr of data of the full-fill phase presented at
Neutrino 2024~\cite{SNO+:nu24, SNO+poster:nu24}.  We show in the right
panel of Fig.~\ref{fig:sun-tension24} the $\Delta\chi^2$ dependence on
$\Dmq_{21}$ after marginalizing over $\theta_{12}$ (fixing
$\sin^2\theta_{13}=0.0222$).  Although the precision of SNO+ is still
far from that of KamLAND, it is interesting to note that its present
best fit is slightly higher than that of KamLAND, though the impact in
the combination is still very marginal as seen in the figure.
Nevertheless, the results from SNO+ and the expected statistics
increase will be interesting to follow due to their potential impact
in the tension/agreement between the solar and reactor determination
of $\Dmq_{21}$.  In that respect, from the figure we read that the
present best-fit value of $\Dmq_{21}$ for the reactor results lies at
$\Delta\chi^2_\text{solar,MB22} = 2.5$, which represent a slight
increase over the $\Delta\chi^2_\text{solar,GS98} = 1.3$ reported in
NuFIT~5.0.  For comparison, we show in orange the results of the solar
analysis without including the Day-Night variation information from
Super-Kamiokande.  As seen in the figure, removing that information
brings the agreement further down to $\Delta\chi^2 \sim 1.5$.
Altogether the latest updates lead to very mild changes in the
determination of ``solar'' parameters ($\sim 1\%$ shift up in the
best-fit value and $\sim 10\%$ improvement in the precision)
reassuring the robustness of the results.

\section{Projections on neutrino mass scale observables}
\label{sec:absmass}

\begin{figure}\centering
  \includegraphics[width=\textwidth]{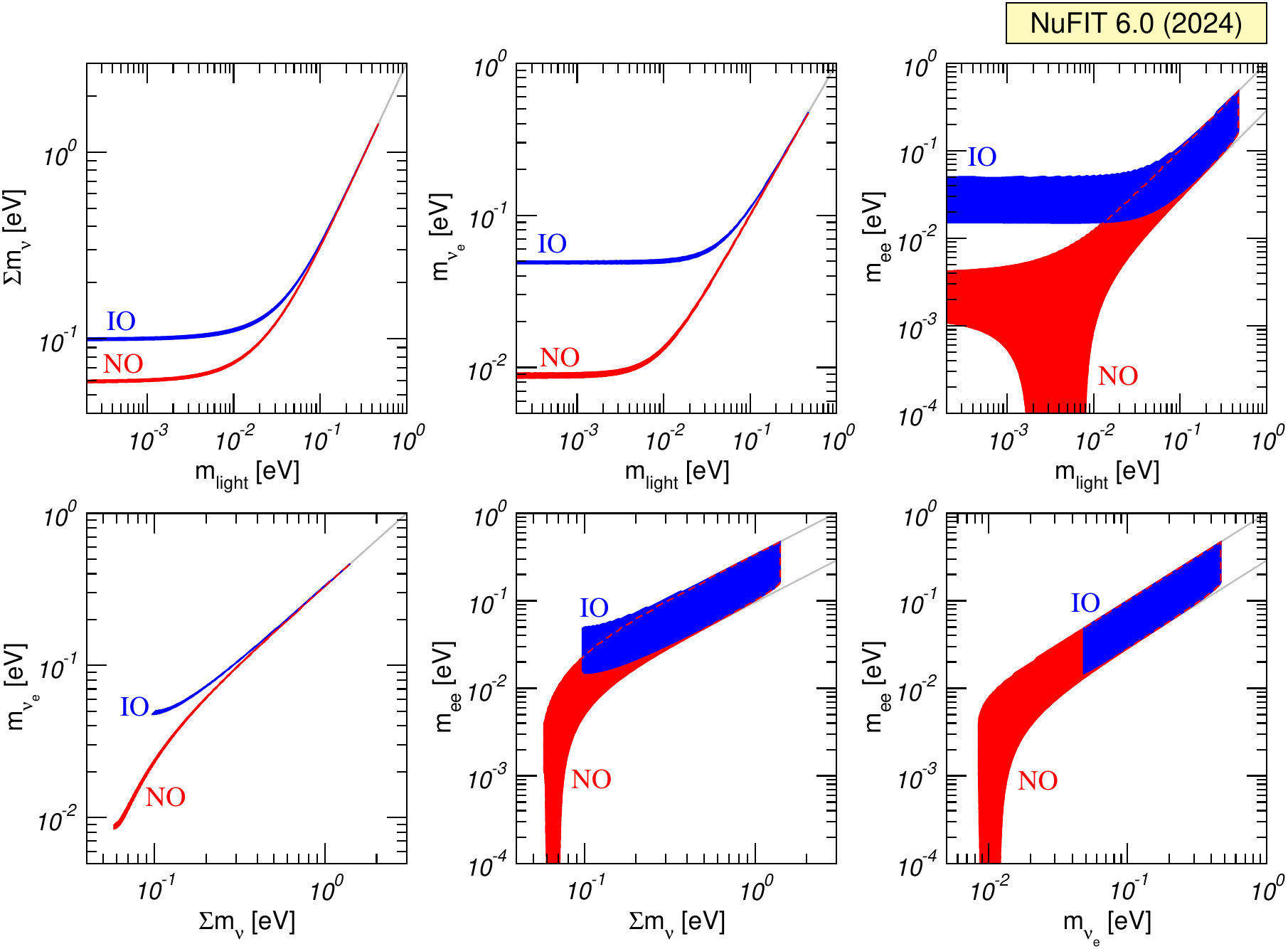}
  \caption{Upper: 95\% CL allowed ranges of the three probes of the
    absolute neutrino mass $\SumNu$, $m_{\nu_e}$, $m_{ee}$ as a
    function of the mass of the lightest neutrino obtained from
    projecting the results of the global analysis of oscillation data.
    The regions are defined with respect to the minimum for each
    ordering.  Lower: Corresponding 95\% CL allowed regions (for 2
    dof) in the planes ($m_{\nu_e}$, $\SumNu$), ($m_{ee}$, $\SumNu$),
    and ($m_{\nu_e}$, $m_{ee}$).}
  \label{fig:mprobes24}
\end{figure}

Because of its quantum-interference nature, mass-induced flavor
oscillations are sensitive to the phase differences induced by the
mass-squared splittings $\Dmq_{ij}$ and to misalignment between the
detection and propagation eigenstates, \textit{i.e.}, to the leptonic
mixing matrix elements $U_{\alpha j}$.  They are, however, insensitive
to overall shifts of the energy levels, and hence they cannot provide
information on the absolute mass scale of the neutrinos other than the
obvious lower bound on the masses of the heaviest states involved in
the oscillations.

The most model-independent information on the neutrino mass, rather
than on mass differences, is obtained from kinematic studies of
reactions in which a neutrino or an antineutrino is involved.  In the
presence of mixing, the most relevant constraint comes from the study
of the end point ($E \sim E_0$) of the electron spectrum in Tritium
beta decay $\Nuc{3}{H} \to \Nuc{3}{He} + e^- + \bar\nu_e$.  This
spectrum can be effectively described by a single parameter,
$m_{\nu_e}$, if for all neutrino states $E_0 - E \gg m_i$:
\begin{equation}
  \label{eq:mbeta}
  m_{\nu_e}^2 \equiv \frac{\sum_i m_i^2 |U_{ei}|^2}{\sum_i |U_{ei}|^2}
  = \sum_i m_i^2 |U_{ei}|^2 \,,
\end{equation}
where the second equality holds if unitarity is assumed.  The most
recent result on the kinematic search for neutrino mass in tritium
decay is from KATRIN~\cite{Katrin:2024tvg}, which sets an upper
limit $m_{\nu_e} < 0.45~\text{eV}$ at 90\% CL.

Direct information on neutrino masses can also be obtained from
neutrinoless double beta decay $(A,Z) \to (A,Z+2) + e^- + e^-$.  This
process violates lepton number by two units, hence in order to induce
the $0\nu\beta\beta$ decay, neutrinos must be Majorana particles.  In
particular, if the only effective lepton number violation at low
energies is induced by a Majorana mass term for neutrinos, the rate of
$0\nu\beta\beta$ decay is proportional to the \emph{effective Majorana
mass of $\nu_e$}:
\begin{equation}
  m_{ee} = \Big| \sum_i m_i U_{ei}^2 \Big| \,.
\end{equation}
Currently the strongest bound on $0\nu\beta\beta$ decay lifetimes are
obtained with Germanium ($T^{0\nu}_{1/2} > 1.8\times 10^{26}$ yr) by
GERDA~\cite{GERDA:2020xhi} and with Xenon ($T^{0\nu}_{1/2} > 3.8\times
10^{26}$ yr) by KamLAND-Zen~\cite{KamLAND-Zen:2024eml}.  Depending on
the assumed nuclear matrix elements, these correspond to 90\% CL
limits of $m_{ee} \lesssim 0.079$--$0.180$~eV~\cite{GERDA:2020xhi} and
$m_{ee} \lesssim 0.028$--$0.122$~eV~\cite{KamLAND-Zen:2024eml},
respectively.

Finally, neutrino masses also have effects in cosmology.  In general,
cosmological data mostly gives information on the sum of the neutrino
masses, $\SumNu$, while it has very little to say on their mixing
structure and on the ordering of the mass states.  At present, no
positive evidence of the cosmological effect of a non-zero neutrino
mass has been observed, which results into upper bounds on $\SumNu$ in
the range of $\SumNu \lesssim 0.04$--$0.3$ eV (see, \textit{e.g.},
Refs.~\cite{Jiang:2024viw, Naredo-Tuero:2024sgf} and references
therein for post-DESI~\cite{DESI:2024mwx} global analyses) depending
on, \textit{e.g.}, the cosmological data included in the analysis,
assumptions on the cosmological model, the statistical approach, the
treatment of systematics, or parameter priors.

Within the $3\nu$-mixing scenario, for each mass ordering, the values
of these observables can be directly predicted in terms of the
parameters determined in the global oscillation analysis and a single
mass scale, which is usually taken to be the lightest neutrino mass
$m_0$.  In addition, the prediction for $m_{ee}$ also depends on the
unknown Majorana phases:
\begin{align}
  m_{\nu_e} & = \sqrt{m_1^2 \, c_{13}^2 c_{12}^2 + m_2^2 \, c_{13}^2 s_{12}^2
  + m_3^2 \, s_{13}^2} \,,
  \\
  m_{ee} &= \left| m_1 \, c_{13}^2 c_{12}^2 \, e^{2i(\alpha_1 - \dCP)}
  + m_2 \, c_{13}^2 s_{12}^2 \, e^{2i(\alpha_2 - \dCP)} + m_3 \, s_{13}^2
  \right| \,,
  \\
  \SumNu &= m_1 + m_2 + m_3 \,,
  \\
  \text{with~}
  &\left\{
  \begin{aligned}
    &\text{NO}:
    &
    m_1 &= m_0 \,,
    &\enspace
    m_2 &= \sqrt{m_0^2 + \Dmq_{21}} \,,
    &\enspace
    m_3 &= \sqrt{m_0^2 + \Dmq_{3\ell}} \,,
    \\
    &\text{IO}:
    &
    m_3 &= m_0 \,
    &\enspace
    m_2 &= \sqrt{m_0^2 - \Dmq_{3\ell}} \,,
    &\enspace
    m_1 &= \sqrt{m_0^2 - \Dmq_{3\ell} - \Dmq_{21}} \,,
  \end{aligned}\right.
\end{align}

We show in the upper panels in Fig.~\ref{fig:mprobes24} the 95\% CL
allowed ranges for these three probes obtained from the projection of
the results of the NuFIT~6.0 global oscillation analysis as a function
of $m_0$.  The larger width of the regions for the $m_{ee}$
predictions is due to the unknown Majorana phases.  The regions are
shown including only the information from the oscillation data (void
regions) and also in combination with the results from KATRIN (filled
regions).  For the latter, we build a $\chi^2$ function trivially
based on their quoted result $m_{\nu_e}^2 = -0.14^{+0.13}_{-0.15}$~eV
under the assumptions of Gaussianity, applicability of Wilks' theorem
(despite the physical boundary $m_{\nu_e}^2\ge 0$), and quadratic
systematics.\footnote{Under these assumptions, $\chi^2(m_{\nu_e}) =
(m_{\nu_e}^2+0.14)^2/0.13^2$ yields a 90\%~CL upper bound
$m_{\nu_e}\leq 0.35$~eV which lies in between the bounds
$m_{\nu_e}\leq 0.45$~eV and $m_{\nu_e}\leq 0.31$~eV obtained by the
collaboration with the Lokhov-Tkachov and Feldman-Cousins methods,
respectively.}

Using the above expressions, one can substitute $m_0$ by any of the
three probes in the expressions of the other two.  Thus, within the
three-neutrino mixing scenario the predicted values for these three
probes are strongly correlated.  We show in the lower panels in
Fig.~\ref{fig:mprobes24} the present status of these correlations.  As
those panels show, with a positive determination of two of these
probes one can in principle obtain information on the value of the
Majorana phases and/or the mass ordering~\cite{Fogli:2004as,
  Pascoli:2005zb}.  Furthermore, a sufficiently strong upper bound can
provide information about the ordering of the
states~\cite{Gariazzo:2022ahe}.

Quantitatively, the global analysis of oscillation data together with
the bound from the KATRIN experiment implies that at 95\% CL
\begin{align}
  0.0085~\text{eV} \leq m_{\nu_e} \leq 0.4~\text{eV}
  &\enspace\text{for NO},
  &\quad
  0.048~\text{eV} \leq m_{\nu_e} \leq 0.4~\text{eV}
  &\enspace\text{for IO},
  \\[2mm]
  0.058~\text{eV} \leq \SumNu \leq 1.2~\text{eV}
  & \enspace\text{for NO},
  &\quad
  0.098~\text{eV} \leq \SumNu \leq 1.2~\text{eV}
  & \enspace\text{for IO}
\end{align}
and for Majorana neutrinos also
\begin{equation}
  0 \leq m_{ee} \leq 0.41~\text{eV}
  \enspace\text{for NO} \,,
  \qquad
  0.015~\text{eV} \leq m_{ee} \leq 0.41~\text{eV}
  \enspace\text{for IO}.
\end{equation}

\section{Summary}
\label{sec:conclu}

We have presented an updated global analysis of world oscillation data
up to September 2024 as listed in Appendix~\ref{sec:appendix-data24}.
Our results are presented in two versions: <<IC19 w/o SK-atm>>
including all the data for which enough information is available to
perform an independent accurate fit, and <<IC24 with SK-atm>>
which includes $\chi^2$ data tables provided by the IceCube and Super
Kamiokande collaborations that we add to our own $\chi^2$.  The global
best-fit values as well as $1\sigma$ and $3\sigma$ ranges for all
parameters are given in Table~\ref{tab:bfranges24}.  The main results
can be summarized as follows:
\begin{itemize}
\item The determination of the parameters $\theta_{12}$,
  $\theta_{13}$, $\Dmq_{21}$, and $|\Dmq_{3\ell}|$ is very stable,
  with Gaussian $\chi^2$ profiles up to high CL.  The relative
  precision at $3\sigma$ for these parameters is about 13\%, 8\%,
  16\%, (5--6)\%, respectively.

\item For $\theta_{23}$ the precision at $3\sigma$ is still about
  20\%, and the determination suffers from the octant ambiguity.
  There is a slight preference for the second octant, $\theta_{23} >
  45^\circ$, (except for NO and the <<IC24 with SK-atm>> data) but for
  all combinations of datasets and mass orderings, the local minimum
  in the other octant always has $\Delta\chi^2 < 4$.

\item The determination of the leptonic CP phase $\dCP$ strongly
  depends on the mass ordering.  For NO the best-fit point is very
  close to the CP-conserving value of $180^\circ$ (with $\Delta\chi^2
  < 1$), and the $\chi^2$ profile is highly non-Gaussian, with some
  dependence on the two data variants and on the octant of
  $\theta_{23}$.  For IO, the best fit points for both data variants
  are close to maximal CP violation $\dCP = 270^\circ$ (within
  $1\sigma$), disfavoring CP conservation at $3.6\sigma$ ($4\sigma$)
  for the <<IC19 w/o SK-atm>> (<<IC24 with SK-atm>>) analysis.

\item Concerning the mass ordering, we find $\Delta\chi^2_\text{IO,NO}
  = -0.6 \,(6.1)$ for the <<IC19 w/o SK-atm>> (<<IC24 with SK-atm>>)
  analysis.  The indecisive result for the <<IC19 w/o SK-atm>>
  analysis emerges from opposite trends in the long-baseline
  accelerator appearance data from T2K and NOvA on the one hand, and
  in the combination of the disappearance channels from accelerator
  and reactor experiments on the other hand.  In the former case, the
  tension between T2K and NOvA for NO has reached $2\sigma$ with the
  latest NOvA update, whereas they are perfectly consistent for IO.
  Conversely, the determination of $|\Dmq_{3\ell}|$ from $\nu_\mu$ and
  $\nu_e$ disappearance agrees better for NO than for IO.  While a
  sensitivity analysis suggests that global <<IC19 w/o SK-atm>> data
  has a median sensitivity of $2.5\sigma$ ($2.8\sigma$) to reject NO
  (IO), the actual result is only $1.7\sigma$ ($1.4\sigma$) for NO
  (IO) because of the opposite trends in the data.  The addition of
  IC24 and SK-atm data provides additional preference for NO leading
  to the above quoted result of $\Delta\chi^2_\text{IO,NO} = 6.1$.
\end{itemize}
We provide also updated ranges and correlations for the effective
parameters sensitive to the absolute neutrino mass from $\beta$-decay,
neutrinoless double-beta decay, and cosmology.  All results and
supplementary material such as additional figures and data tables are
provided at the NuFit webpage~\cite{nufit}.

\acknowledgments

We would like to thank Tetiana Kozynets and Philipp Eller for useful
discussions about the IceCube simulation.  This project is funded by
USA-NSF grant PHY-2210533 and by the European Union's through the
Horizon 2020 research and innovation program (Marie
Sk{\l}odowska-Curie grant agreement 860881-HIDDeN) and the Horizon
Europe research and innovation programme (Marie Sk{\l}odowska-Curie
Staff Exchange grant agreement 101086085-ASYMMETRY), and by ERDF ``A
way of making Europe''.  It also receives support from grants
PID2022-\allowbreak 126224NB-\allowbreak C21, PID2022-\allowbreak
136510NB-\allowbreak C33, PID2022-\allowbreak 142545NB-\allowbreak
C21, PID2021-\allowbreak 123703NB-\allowbreak C21, ``Unit of
Excellence Maria de Maeztu 2020-2023'' award to the ICC-UB
CEX2019-000918-M, grant IFT ``Centro de Excelencia Severo Ochoa''
CEX2020-001007-S funded by MCIN/AEI/\allowbreak 10.13039/\allowbreak
501100011033, as well as from grants 2021-SGR-249 (Generalitat de
Catalunya) and from Basque Government (IT1628-22) grant.  IMS is
supported by the STFC under Grant No.~ST/X003167/1.  Part of this work
used the Solaris cluster, acquired through the Basque Government
IT1628-22 grant.

\appendix

\section{List of data used in the analysis}
\label{sec:appendix-data24}

\section*{Solar experiments}

\begin{itemize}
\item[\NEW] \emph{External information}: Standard Solar
  Models~\cite{B23Fluxes}.

\item Chlorine total rate~\cite{Cleveland:1998nv}, 1 data point.

\item Gallex \& GNO total rates~\cite{Kaether:2010ag}, 2 data points.

\item SAGE total rate~\cite{Abdurashitov:2009tn}, 1 data point.

\item SK1 1496-day energy and zenith spectrum~\cite{Hosaka:2005um}, 44
  data points.

\item SK2 791-day energy and day/night spectrum~\cite{Cravens:2008aa},
  33 data points.

\item SK3 548-day energy and day/night spectrum~\cite{Abe:2010hy}, 42
  data points.

\item[\NEW] SK4 2970-day energy and day/night
  spectrum~\cite{Super-Kamiokande:2023jbt}, 46 data points.

\item SNO combined analysis~\cite{Aharmim:2011vm}, 7 data points.

\item Borexino Phase-I 741-day low-energy data~\cite{Bellini:2011rx},
  33 data points.

\item Borexino Phase-I 246-day high-energy data~\cite{Bellini:2008mr},
  6 data points.

\item[\NEW] Borexino Phase-II 1292-day low-energy
  data~\cite{Borexino:2017rsf}, 192 data points.

\item[\NEW] Borexino Phase-III 1432-day low-energy
  data~\cite{BOREXINO:2022abl}, 120 data points.
\end{itemize}

\section*{Atmospheric experiments}

\begin{itemize}
\item \emph{External information}: Atmospheric neutrino
  fluxes~\cite{Honda:2015fha}.

\item[\NEW] IC19 IceCube/DeepCore 3-year data
  (2012-2015)~\cite{IceCube:2019dqi, IceCube:2019dyb}, 140 data points.

\item[\NEW] {IC24 IceCube/DeepCore 9.3-year data (2012-2021)
  $\chi^2$   map~\cite{IceCube:2024xjj, IC:data2024}
  added to our global analysis.}

\item[\NEW] SK1-5 484.2 kiloton-year
  data~\cite{Super-Kamiokande:2023ahc}, $\chi^2$
  map~\cite{SKatm:data2024} added to our global analysis.

\end{itemize}

\section*{Reactor experiments}

\begin{itemize}
\item[\NEW] KamLAND separate DS1, DS2, DS3 spectra~\cite{Gando:2013nba} with
  Daya Bay reactor $\nu$ fluxes~\cite{DayaBay:2021dqj}, 69 data points.

\item[\NEW] SNO+ spectrum from
  partial fill 114 ton-yr~\cite{SNO:2024wzq} data and full fill 286
  ton-yr data~\cite{SNO+:nu24, SNO+poster:nu24}, 17 data points.

\item Double-Chooz FD/ND spectral ratio, with 1276-day (FD), 587-day
  (ND) exposures~\cite{DoubleC:nu2020}, 26 data points.

\item[\NEW] Daya Bay 3158-day separate EH1, EH2, EH3
  spectra~\cite{DayaBay:2022orm}, 78 data points.

\item Reno 2908-day FD/ND spectral ratio~\cite{RENO:nu2020}, 45 data
  points.
\end{itemize}

\section*{Accelerator experiments}

\begin{itemize}
\item MINOS $10.71\times 10^{20}$~pot $\nu_\mu$-disappearance
  data~\cite{Adamson:2013whj}, 39 data points.

\item MINOS $3.36\times 10^{20}$~pot $\bar\nu_\mu$-disappearance
  data~\cite{Adamson:2013whj}, 14 data points.

\item MINOS $10.6\times 10^{20}$~pot $\nu_e$-appearance
  data~\cite{Adamson:2013ue}, 5 data points.

\item MINOS $3.3\times 10^{20}$~pot $\bar\nu_e$-appearance
  data~\cite{Adamson:2013ue}, 5 data points.

\item[\NEW] T2K $21.4\times 10^{20}$ pot $\nu_\mu$-disappearance
  data~\cite{T2K:nu24}, 28 data points.

\item[\NEW] T2K $21.4\times 10^{20}$ pot $\nu_e$-appearance
  data~\cite{T2K:nu24}, 9 data points for the CCQE and 7 data
  points for the CC1$\pi$ samples.

\item[\NEW] T2K $16.3\times 10^{20}$ pot $\bar\nu_\mu$-disappearance
  data~\cite{T2K:2023mcm}, 19 data points.

\item[\NEW] T2K $16.3\times 10^{20}$ pot $\bar\nu_e$-appearance
  data~\cite{T2K:2023smv}, 9 data points.

\item[\NEW] NOvA $26.6\times 10^{20}$ pot $\nu_\mu$-disappearance
  data~\cite{NOvA:nu24}, 22 data points.

\item[\NEW] NOvA $26.6\times 10^{20}$ pot $\nu_e$-appearance
  data~\cite{NOvA:nu24}, 15 data points.

\item NOvA $12.5\times 10^{20}$ pot $\bar{\nu}_\mu$-disappearance
  data~\cite{NOvA:nu2020}, 76 data points.

\item NOvA $12.5\times 10^{20}$ pot $\bar{\nu}_e$-appearance
  data~\cite{NOvA:nu2020}, 13 data points.
\end{itemize}

\section{IceCube 2019}
\label{sec:app_IC}

The IceCube analysis (IC19) is based on Analysis A from
Refs.~\cite{IceCube:2019dqi, IceCube:2019dyb}.  This analysis uses
data collected over 3-years, from April 2012 to May 2015.  Following
the public data release~\cite{Collaboration_2019}, we computed the
expected number of events for each bin, where the reconstructed energy
is logarithmically distributed between $5.6$~GeV and $56$~GeV across
eight bins, and the reconstructed cosine of zenith angle is
distributed between $-1$ and $+1$ across ten bins.  The events are
categorized into two particle identification (PID) types: tracks and
cascades.  The expected number of events in each bin is given by
\begin{equation}
  \label{eq:Ev}
    N_{i} = T\sum_{\alpha}\sum_{\text{MC}}
    \phi_{\alpha}^{\text{atm}}(E_{\nu},\cos\theta) \,
    P_{\alpha\beta}(E_{\nu},\cos\theta) \, A^\text{eff}_{\beta}
\end{equation}
where the sum is performed over all the Monte Carlo simulated events
that contribute to the $i$-th bin and over all initial flavor states,
corresponding to electron and muon neutrinos in the case of
atmospheric neutrinos.  For the atmospheric neutrino flux, we used the
Honda flux tables~\cite{Honda:2015fha} with the energy spectra
modified by a factor of $(E_{\nu} \big/ 5~\text{GeV})^{-0.05}$.  Both
the flux and the oscillation probabilities are evaluated at the true
energy and zenith of the simulated events.  The constant $T$ denotes
the total data-taking time.  In addition to the expected event
distribution, each bin includes a contribution from the atmospheric
muon background, determined by binning the simulated muon background
events provided in the public release.

The event distribution in each bin is modified by the detector
systematics, which account for the uncertainties in the detector
response.  These uncertainties include factors such as optical
absorption, photon scattering in the ice, DOM efficiency, and the
coincidence in reconstruction between neutrinos and atmospheric muons.
The systematics are applied to the event distribution as
multiplicative reweighting factors for each bin.  We have incorporated
these uncertainties into our analysis using the information provided
in the data release and taking the variable called \verb"reco_coszen"
as the negative of the reconstructed zenith angle
($-\cos\theta_{rec}$).  These systematics apply to both the event and
background distributions.  For the uncertainties, we used the values
specified in the data release.  For the nominal prediction, we assumed
that the event distribution is corrected by the offset plus $5 \times
\verb"opt_eff_lateral"$.

In addition to detector-related systematics, the analysis incorporates
uncertainties associated with the atmospheric neutrino flux.  These
uncertainties are accounted for as follows:
\begin{itemize}
\item Normalization: An uncertainty of 100\% was assumed, although the
  results remain consistent when this parameter is left free.

\item Initial flavor composition: A 18\% uncertainty was included to
  account for variations in the initial composition of electron and
  muon neutrinos.

\item Neutrino to antineutrino ratio: An uncertainty of 25\% was
  applied to account for variations in the relative flux of neutrinos
  and antineutrinos.

\item Energy dependence of the flux: This uncertainty is parameterized
  using the factor $(E_{\nu}/E_{0})^{\gamma}$, with $E_{0}=5$~GeV.
  The central value of $\gamma$ is set to zero, and an uncertainty of
  5\% is applied.

\item Upward vs.\ horizontal flux ratio: A 10\% uncertainty was
  assumed to account for directional asymmetries in the atmospheric
  neutrino flux.
\end{itemize}

\begin{figure}\centering
  \includegraphics[width=1\textwidth]{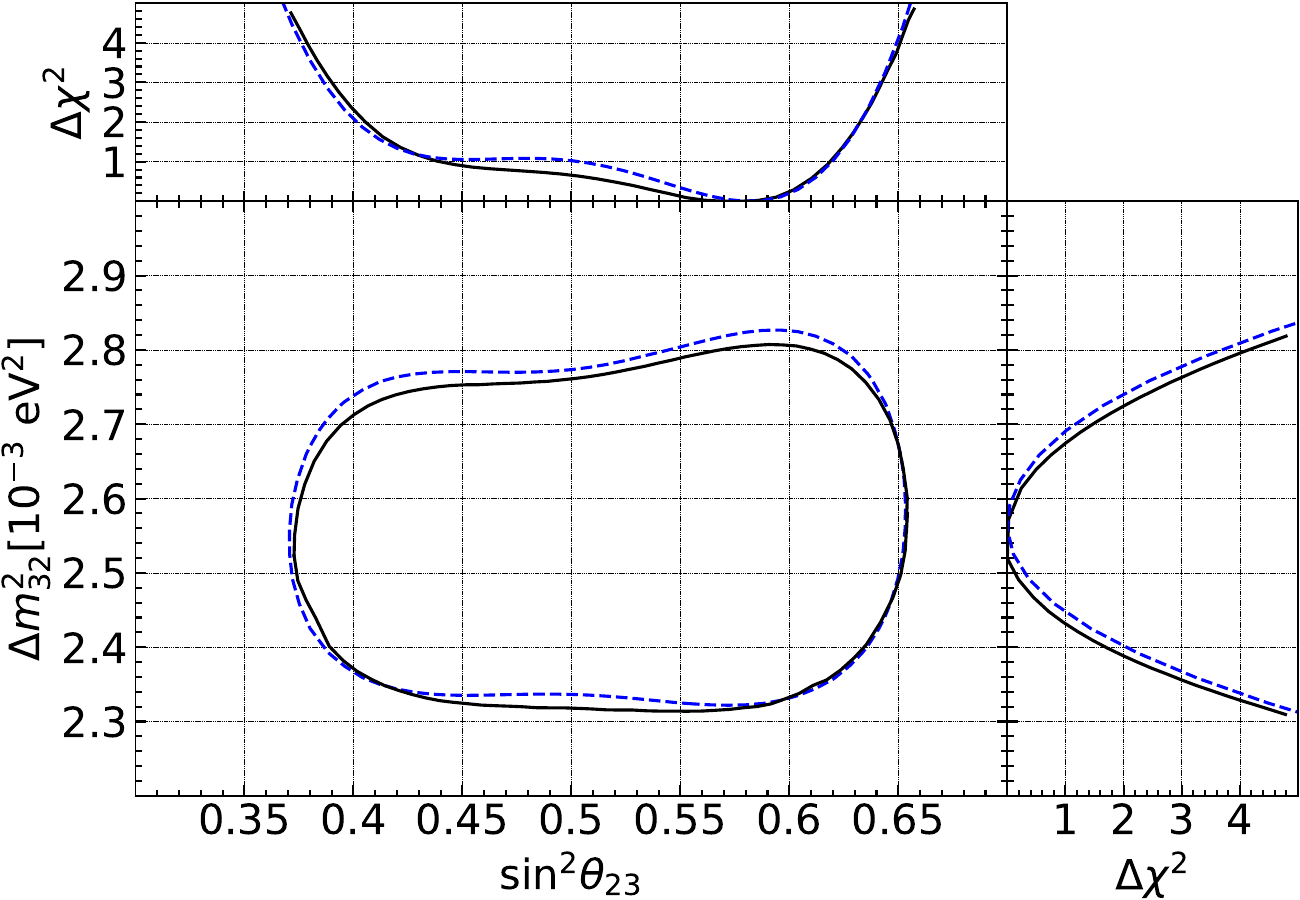}
    \caption{Our fit to IceCube Analysis A (blue dashed lines) is
      compared to the results from IceCube~\recite{IceCube:2019dqi}
      (black lines).
      %We have also estimated the sensitivity to the mass ordering, finding a preference for normal ordering (NO) with a difference of 0.7 units in $\chi^2$.
    }
\label{fig:IC19}
\end{figure}

Assuming a Gaussian $\chi^2$ distribution and noting that the variable
\verb"reco_coszen" in the data corresponds to the negative of
\verb"reco_coszen" in the simulated events, we have computed IceCube’s
sensitivity to $\Dmq_{32}$ and $\sin^2\theta_{23}$.  The results are
presented in Fig.~\ref{fig:IC19}, alongside a comparison with the
findings from Analysis A of~\cite{IceCube:2019dqi}.  By comparing the
best-fit points for the two neutrino mass orderings we obtain a
non-significant preference for the normal ordering with a difference
of $0.7$ units in $\chi^2$, in excellent agreement with the
corresponding result $0.738$ reported in ref.~\cite{IceCube:2019dyb}.

\section{Assumed true values for the MO test}
\label{sec:app-T}

As mentioned in Sec.~\ref{sec:MO}, the values of $T_0^o$ ---~and
therefore the distribution of $T$~--- depend on the unknown true value
of the oscillation parameters, see Eq.~\eqref{eq:T0def}.  In
principle, one needs to consider the distribution of $T$ for all
possible values of $\theta^\text{true}$ and the final $p$-value of a
MO hypothesis will be given by the largest one among all choices of
$\theta^\text{true}$, \textit{i.e.}, by the weakest rejection (see the
discussion in Ref.~\cite{Blennow:2013oma}).  In the main text, we have
assumed that the best fit of the real data is representative of the
$T$-distribution at the unknown true value of $\theta$.  Indeed, given
the allowed regions of the oscillation parameters, we do not expect
$T_0$ to change significantly if we vary $\theta^\text{true}$ within
the allowed regions at reasonable confidence level.  The only
exception may be the sensitivity of LBL data as a function of $\dCP$,
which is known to affect the MO sensitivity, especially for NOvA as
shown in Fig.~\ref{fig:nevts24}.

For IO, the global fit constrains $\dCP$ reasonably well, so that we
do not expect strong variations of $T_0^\text{IO}$ for true values of
$\dCP$ in its allowed range for IO.  However, for NO, a significantly
larger range of $\dCP$ is allowed, in particular in correlation with
$\theta_{23}$, see Fig.~\ref{fig:region-cp23}.  Therefore, the
question arises, whether the MO test would give largely different
results when considering true values for $\dCP$ and $\theta_{23}$
within the allowed region for NO.

From Fig.~\ref{fig:nevts24}, we see that for $\dCP \approx 270^\circ$
and NO, the number of $\nu_e$ events in NOvA is maximal, and its value
cannot be obtained by any parameter choice in IO.  Therefore, we
expect best sensitivity to NO for this value of $\dCP$.  In contrast,
values around 0 or $180^\circ$ can be easily accommodated within IO,
and we expect that the sensitivity to NO is weakest around
CP-conserving values.  We have confirmed this expectation, as we
obtain
\begin{equation}
  \label{eq:T0max}
  T_0^\text{NO}\big( \dCP^\text{true}=270^\circ,
  \sin^2\theta_{23}^\text{true} = 0.46 \big) = 12.38 \,,
\end{equation}
with all other oscillation parameters kept at their best-fit values.
This value is significantly larger than the value in Eq.~\eqref{eq:T0}
for the best-fit point with $\dCP = 177^\circ$, implying higher
sensitivity to reject NO.  Actually, for the value in
Eq.~\eqref{eq:T0max} the observed value $T_\text{obs} = -0.6$ would
imply a $p$-value for NO of 3.2\%.  Since values of $\dCP$ around
$90^\circ$ are significantly disfavored also for NO, we do not
consider them relevant for the MO test.

To summarize, for relevant choices of the oscillation parameters, the
sensitivity to the NO is weakest for values of $\dCP$ around
$180^\circ$.  In turn, for IO, $\dCP$ is sufficiently constrained, and
we expect only minor variations of $T_0$ within the relevant range
around $270^\circ$.  Therefore, it is appropriate to consider the
current best fit points to quote the final $p$-values for both
orderings.

\bibliographystyle{JHEPmod}
\bibliography{references}

\end{document}